\documentclass[twocolumn]{aastex63}
\usepackage{url}
\usepackage{threeparttable}
\usepackage{ulem}

\usepackage{natbib}
\usepackage{appendix}
\usepackage{multirow}
\usepackage{longtable}
\usepackage{tabularx}
\usepackage{ltablex}
\keepXColumns

\def\tcra{\textcolor{black}}
\def\tcrb{\textcolor{black}}

\def\hii{H\,\textsc{ii}}
\def\oii{O\,\textsc{ii}}
\def\oiii{O\,\textsc{iii}}
\def\sii{S\,\textsc{ii}}
\def\siii{S\,\textsc{iii}}
\def\ciii{C\,\textsc{iii}}

\def\heii{He\,\textsc{ii}}

\def\niii{N\,\textsc{iii}}
\def\niv{N\,\textsc{iv}}

\def\neiii{Ne\,\textsc{iii}}
\def\neiv{Ne\,\textsc{iv}}

\def\ariii{Ar\,\textsc{iii}}

\newcommand{\approptoinn}[2]{\mathrel{\vcenter{
  \offinterlineskip\halign{\hfil$##$\cr
    #1\propto\cr\noalign{\kern2pt}#1\sim\cr\noalign{\kern-2pt}}}}}

\accepted{for publication in ApJ}
\shorttitle{
Extremely Low C/N Galaxies with [N/O]$\gtrsim 0.5$ at $z\sim 6-10$
}

\shortauthors{Isobe et al.}
\graphicspath{{./}{figures/}}

\begin{document}

\title{
%
%
%
%
JWST Identification of Extremely Low C/N Galaxies with [N/O]$\gtrsim 0.5$ at $z\sim 6-10$\\
Evidencing the Early CNO-Cycle Enrichment and a Connection with Globular Cluster Formation
}

\author[0000-0001-7730-8634]{Yuki Isobe}
\affiliation{Institute for Cosmic Ray Research, The University of Tokyo, 5-1-5 Kashiwanoha, Kashiwa, Chiba 277-8582, Japan}
\affiliation{Department of Physics, Graduate School of Science, The University of Tokyo, 7-3-1 Hongo, Bunkyo, Tokyo 113-0033, Japan}
\email{isobe@icrr.u-tokyo.ac.jp}

\author[0000-0002-1049-6658]{Masami Ouchi}
\affiliation{National Astronomical Observatory of Japan, 2-21-1 Osawa, Mitaka, Tokyo 181-8588, Japan}
\affiliation{Institute for Cosmic Ray Research, The University of Tokyo, 5-1-5 Kashiwanoha, Kashiwa, Chiba 277-8582, Japan}
\affiliation{Department of Astronomical Science, SOKENDAI (The Graduate University for Advanced Studies), Osawa 2-21-1, Mitaka, Tokyo, 181-8588, Japan}
\affiliation{Kavli Institute for the Physics and Mathematics of the Universe (WPI), University of Tokyo, Kashiwa, Chiba 277-8583, Japan}

\author[0000-0001-8537-3153]{Nozomu Tominaga}
\affiliation{National Astronomical Observatory of Japan, 2-21-1 Osawa, Mitaka, Tokyo 181-8588, Japan}
\affiliation{Department of Physics, Faculty of Science and Engineering, Konan University, 8-9-1 Okamoto, Kobe, Hyogo 658-8501, Japan}
\affiliation{Department of Astronomical Science, SOKENDAI (The Graduate University for Advanced Studies), Osawa 2-21-1, Mitaka, Tokyo, 181-8588, Japan}

\author{Kuria Watanabe}
\affiliation{Department of Astronomical Science, SOKENDAI (The Graduate University for Advanced Studies), Osawa 2-21-1, Mitaka, Tokyo, 181-8588, Japan}

\author[0000-0003-2965-5070]{Kimihiko Nakajima}
\affiliation{National Astronomical Observatory of Japan, 2-21-1 Osawa, Mitaka, Tokyo 181-8588, Japan}

\author[0009-0008-0167-5129]{Hiroya Umeda}
\affiliation{Institute for Cosmic Ray Research, The University of Tokyo, 5-1-5 Kashiwanoha, Kashiwa, Chiba 277-8582, Japan}
\affiliation{Department of Physics, Graduate School of Science, The University of Tokyo, 7-3-1 Hongo, Bunkyo, Tokyo 113-0033, Japan}

\author[0000-0002-1319-3433]{Hidenobu Yajima}
\affiliation{Center for Computational Sciences, University of Tsukuba, Ten-nodai, 1-1-1 Tsukuba, Ibaraki 305-8577, Japan}

\author[0000-0002-6047-430X]{Yuichi Harikane} 
\affiliation{Institute for Cosmic Ray Research, The University of Tokyo, 5-1-5 Kashiwanoha, Kashiwa, Chiba 277-8582, Japan}

\author[0000-0002-0547-3208]{Hajime Fukushima}
\affiliation{Center for Computational Sciences, University of Tsukuba, Ten-nodai, 1-1-1 Tsukuba, Ibaraki 305-8577, Japan}

\author[0000-0002-5768-8235]{Yi Xu}
\affiliation{Institute for Cosmic Ray Research, The University of Tokyo, 5-1-5 Kashiwanoha, Kashiwa, Chiba 277-8582, Japan}
\affiliation{Department of Astronomy, Graduate School of Science, The University of Tokyo, 7-3-1 Hongo, Bunkyo, Tokyo 113-0033, Japan}

\author[0000-0001-9011-7605]{Yoshiaki Ono}
\affiliation{Institute for Cosmic Ray Research, The University of Tokyo, 5-1-5 Kashiwanoha, Kashiwa, Chiba 277-8582, Japan}

\author[0000-0003-3817-8739]{Yechi Zhang}
\affiliation{Institute for Cosmic Ray Research, The University of Tokyo, 5-1-5 Kashiwanoha, Kashiwa, Chiba 277-8582, Japan}
\affiliation{Department of Astronomy, Graduate School of Science, The University of Tokyo, 7-3-1 Hongo, Bunkyo, Tokyo 113-0033, Japan}




\begin{abstract}
We present chemical abundance ratios of 70 star-forming galaxies at $z\sim4$--10 observed by the JWST/NIRSpec ERO, GLASS, and CEERS programs.
Among the 70 galaxies, we have pinpointed 2 galaxies, CEERS\_01019 at $z=8.68$ and GLASS\_150008 at $z=6.23$, with extremely low C/N ([C/N]$\lesssim -1$), evidenced with
\ciii]$\lambda\lambda$1907,1909, \niii]$\lambda$1750, and \niv]$\lambda\lambda$1483,1486, which show high N/O ratios ([N/O]$\gtrsim 0.5$) comparable with the one of GN-z11 \tcra{regardless of whether stellar or AGN radiation is assumed}.
Such low C/N and high N/O ratios found in CEERS\_01019 and GLASS\_150008 (additionally identified in GN-z11) are \tcrb{largely biased towards} the equilibrium of the CNO cycle, suggesting that these 3 galaxies are enriched by metals processed by the CNO cycle.
On the C/N vs. O/H plane, these 3 galaxies do not coincide with Galactic {\sc Hii} regions, normal star-forming galaxies, and nitrogen-loud quasars with \tcra{asymptotic giant branch} stars, but globular-cluster (GC) stars, indicating a connection with GC formation.
We compare C/O and N/O of these 3 galaxies with those of theoretical models, and find that these 3 galaxies are explained by scenarios with dominant CNO-cycle materials,
i.e. Wolf-Rayet stars, supermassive ($10^{3}-10^{5}\ M_{\odot}$) stars, and tidal disruption events, interestingly with a requirement of frequent direct collapses. For all the 70 galaxies, we present measurements of Ne/O, S/O, and Ar/O, together with C/O and N/O. We identify 4 galaxies with very low Ne/O, $\log(\rm Ne/O)<-1.0$, indicating abundant massive ($\gtrsim30\ M_\odot$) stars.
\end{abstract}

\keywords{High-redshift galaxies (734); Galaxy chemical evolution (580); Galaxy formation (595); Star formation (1569)}


\section{Introduction} \label{sec:intro}
Chemical abundance ratios of the inter-stellar medium (ISM) in early galaxies are crucial to understanding stellar nucleosynthesis.
Local dwarf star-forming galaxies have $\alpha$-to-oxygen ($\alpha$/O) ratios such as neon-to-oxygen (Ne/O) that are around solar abundances and remain mostly constant for their gas-phase metallicity \citep[e.g.,][]{Izotov2006,Kojima2021,Isobe2022}.
These findings suggest that the majority of $\alpha$ elements are produced by massive stars evolving into core-collapse supernovae (CCSNe) and/or hypernovae (HNe; e.g., \citealt{Nomoto2013}).
On the other hand, nitrogen-to-oxygen (N/O) ratios of star-forming galaxies increase with metallicity \citep[e.g.,][]{Izotov2006,Pilyugin2012,Kojima2021,Isobe2022}, thought to originate from primary oxygen production by massive stars and secondary nitrogen production by low and intermediate-mass metal-rich stars evolving to asymptotic giant branch (AGB) stars \citep[e.g.,][]{Vincenzo2016}.
Similarly, carbon-to-oxygen (C/O) ratios of star-forming galaxies increase with metallicity due to massive stars and AGB stars \citep[e.g.,][]{Berg2019}.

Before the arrival of the James Webb Space Telescope (JWST), these kinds of chemical abundance studies were restricted to at most intermediate redshifts, even with stacking analysis \citep{Steidel2016} or lensed objects \citep[e.g.,][]{Kojima2017}.
However, Near Infrared Spectrograph (NIRSpec) on JWST can spectroscopically observe a near-infrared (NIR) wavelength range of 1--5 $\mu$m 10--1000 times more deeply than other spectrographs.
This great advancement has led to several emission line detections of high-redshift ($z\gtrsim4$) galaxies in restframe ultraviolet (UV) to optical ranges, which are vital for chemical abundance measurements.
Shortly after Early Release Observations (ERO; \citealt{Pontoppidan2022}; PID: 2736) became public, \citet{Arellano2022} have reported Ne/O ratios of 3 ERO galaxies and the C/O ratio of one ERO galaxy at $z>7$.
The 3 and 1 galaxies have Ne/O and C/O ratios comparable to those of local galaxies at a given metallicity, respectively.

However, \citet{Jones2023arXiv} have found that a $z=6.23$ galaxy observed by the GLASS-JWST Early Release Science (ERS) program (hereafter GLASS; \citealt{Treu2022}; PID: 1324) has a very low $\log(\rm C/O)$ value of $-1.01\pm0.22$.
As the low C/O value is consistent with those expected from yields of pure core-collapse supernovae \citep{Nomoto2013}, \citet{Jones2023arXiv} have interpreted the galaxy as having a young chemical composition where carbon production by AGB stars is negligible.

Moreover, GN-z11 at $z=10.6$, photometricly identified by \citet{Oesch2016} and spectroscopically confirmed by the JWST Advanced Deep Extragalactic Survey (JADES; \citealt{Bunker2023arXiv}), has a super-solar value of $\log(\rm N/O)\gtrsim-0.5$ \citep{Cameron2023btmp,Senchyna2023arXiv}, which is significantly higher than those of local galaxies at a given metallicity.
Given the ultra-high redshift of $z=10.6$ corresponding to 440 Myr after the Big Bang, nitrogen production by AGB stars is difficult to explain the N/O enhancement of GN-z11 \citep[e.g.,][]{Cameron2023btmp}.
Since the report, several studies have been trying to explain the high N/O ratio of GN-z11 (see Section \ref{subsec:nco_org} for more details).
However, it remains unknown whether there is another galaxy with super-solar N/O ratios at high redshift.
The carbon-nitrogen (C/N) ratio has also not been studied, although it is usually assumed that both carbon and nitrogen originate from AGB stars.

Furthermore, given that chemical evolution models \citep{Watanabe2023arXiv} predict significant changes of Ne/O ratios from the solar abundances at the early formation phase (more details in Section \ref{subsec:neo_dis}), Ne/O ratios potentially evolve toward higher redshifts.

The aim of this paper is to study the neon, carbon, nitrogen, and oxygen abundances of high-$z$ galaxies using data from the NIRSpec public surveys to characterize the nature of star formation at high redshifts.
This paper is organized as follows.
Section \ref{sec:datsam} explains the data and sample we use.
Our analysis is described in Section \ref{sec:analysis}.
We report and discuss our results of nitrogen, carbon, and oxygen abundances in Sections \ref{sec:nco}.
Results and discussions of Ne/O ratios are described in \ref{sec:neo}.
Our findings are summarized in Section \ref{sec:sum}.
We assume a standard $\Lambda$CDM cosmology with parameters of ($\Omega_{\rm m}$, $\Omega_{\rm \Lambda}$, $H_{0}$) = (0.3, 0.7, 70 km ${\rm s}^{-1}$ ${\rm Mpc}^{-1}$). 
Throughout this paper, we use the solar abundance ratios of \citet{Asplund2021}.
The notation [A/B] is defined as $\log(\rm A/B)$ subtracted by the solar abundance $\log(\rm A/B)_{\odot}$.

\section{Data and Sample} \label{sec:datsam}
\begin{figure*}[t]
    \centering
    \includegraphics[width=18.0cm]{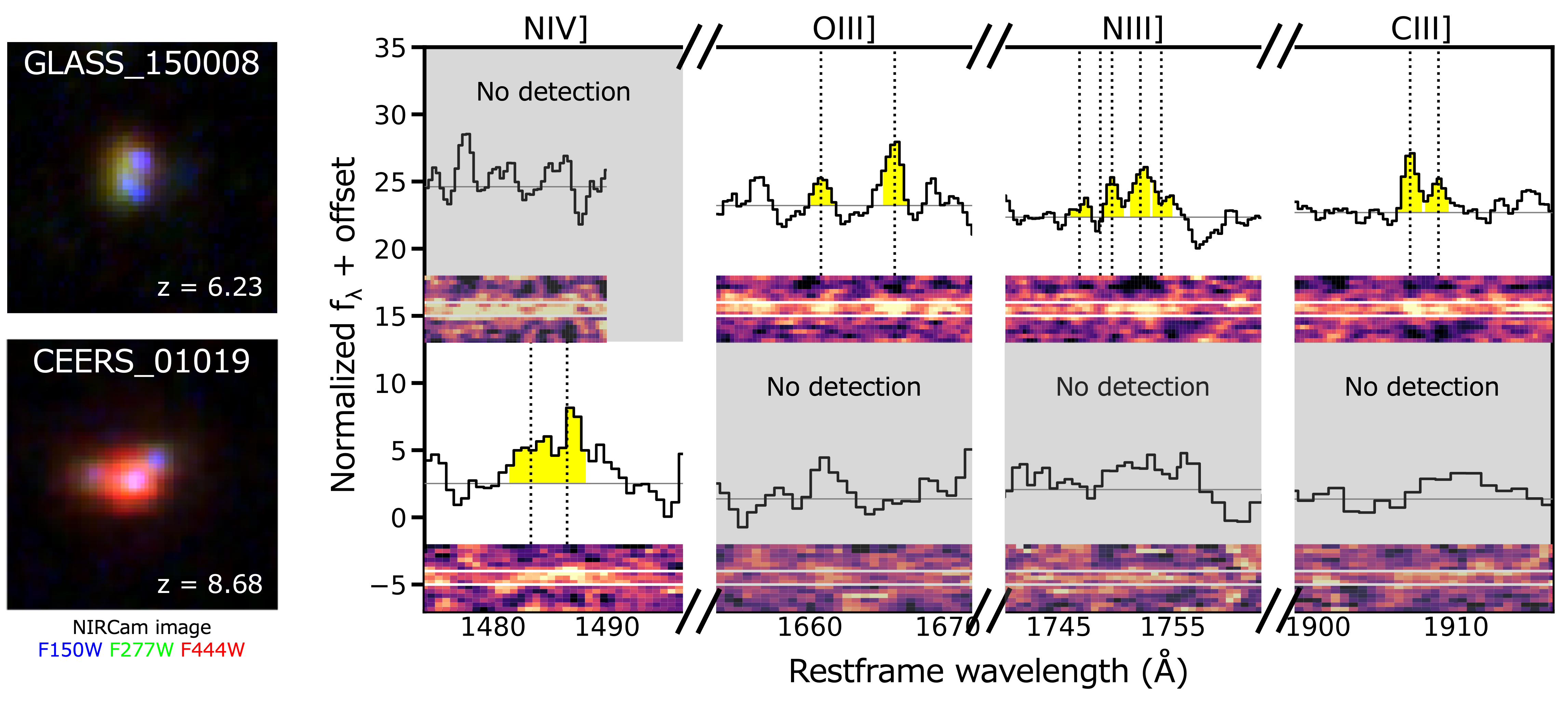}
    \caption{\tcra{Emission line detections of GLASS\_150008 and CEERS\_01019.
    (Left) $1^{\prime\prime}\times1^{\prime\prime}$ JWST/NIRCam image with blue/green/red corresponding to F150W/F277W/F444W.
    (Right) Smoothed 1D and 2D spectra around \niv], \oiii], \niii], and \ciii] shown by the black solid lines and the color maps, respectively. The emission lines are highlighted in yellow, while the highlights for the undetected emission lines are omitted. Restframe wavelengths of the emission lines are shown by the vertical black dotted lines. The horizontal gray lines denote the continuum level. The horizontal white lines on the 2D spectra indicate the extraction aperture for the 1D spectra.}}
    \label{fig:spec}
\end{figure*}

We utilize JWST/NIRSpec data from the Early Release Observations (ERO; \citealt{Pontoppidan2022}), taken in the SMACS 0723 lensing cluster field (hereafter referred to as ERO data), the GLASS survey (\citealt{Treu2022}; hereafter referred to as GLASS data), and the CEERS survey (\citealt{Finkelstein2022arXiv}; hereafter referred to as CEERS data). The ERO data were acquired using medium-resolution ($R\sim1000$) filter-grating pairs of F170LP-G235M and F290LP-G395M, covering the wavelength ranges of 1.7--3.1 and 2.9--5.1 $\mu$m, respectively. The total exposure time for the ERO data is 4.86 hours for each filter-grating pair. The GLASS data were collected using high-resolution ($R\sim2700$) filter-grating pairs of F100LP-G140H, F170LP-G235H, and F290LP-G395H, covering the wavelength ranges of 1.0--1.6, 1.7--3.1, and 2.9--5.1 $\mu$m, respectively. The total exposure time for the GLASS data is 4.9 hours for each filter-grating pair. The CEERS data were obtained using medium-resolution filter-grating pairs of F100LP-G140M, F170LP-G235M, and F290LP-G395M, covering the wavelength ranges of 1.0--1.6, 1.7--3.1, and 2.9--5.1 $\mu$m, respectively. The total exposure time for the CEERS data is 0.86 hours for each filter-grating pair.

We use spectroscopic data that has been reduced by \citet{Nakajima2023}. \citet{Nakajima2023} have extracted the raw data from the MAST archive and performed level-2 and 3 calibrations using the JWST Science Calibration Pipeline \tcra{(ver.1.8.5; \citealt{JWSTpipeline185})} with the reference file jwst\_1028.pmap, whose flux calibration is based on in-flight flat data.
Checking the data, we identify 5, 15, and 50 galaxies at $z>4$ in the ERO, GLASS, and CEERS data, respectively.
We refer to these 70 ($=5+15+50$) galaxies as our sample galaxies, hereafter.

\section{Analysis} \label{sec:analysis}
\subsection{Flux Measurement} \label{subsec:flux}
We measure emission-line fluxes by fitting a Gaussian function convolved by line-spread functions that \citet{Isobe2023} have derived from in-flight NIRSpec data of a planetary nebula (PI: J. Muzerolle; PID: 1125).
To measure weak emission lines accurately, we fix redshift and velocity dispersion of the Gaussian function to those of [\oiii]$\lambda\lambda$4959,5007.
To derive Ne/O, C/O, and N/O ratios, we explore the following line detections: [\neiii]$\lambda$3869 (for neon abundance), \ciii]$\lambda\lambda$1907,1909 (for carbon abundance), \niii]$\lambda$1750\footnote{We refer to the quintet of \niii] lines at restframe 1747, 1749, 1750, 1752, and 1754 \AA\ as \niii]$\lambda$1750 in this paper.} and \niv]$\lambda\lambda$1483,1486 (for nitrogen abundance), [\oii]$\lambda\lambda$3727,3729, [\oiii]$\lambda$5007, and \oiii]$\lambda\lambda$1661,1666 (for oxygen abundance), [\oiii]$\lambda$4363 (for electron temperature $T_{\rm e}$), and Balmer lines of H$\alpha$, H$\beta$, and H$\gamma$ (for hydrogen abundance and dust correction).
Hereafter, we just write [\neiii], \ciii], \niii], \niv], [\oii], [\oiii], and \oiii] as meaning [\neiii]$\lambda$3869, \ciii]$\lambda\lambda$1907,1909, \niii]$\lambda$1750, \niv]$\lambda\lambda$1483,1486, [\oii]$\lambda\lambda$3727,3729, [\oiii]$\lambda$5007, and \oiii]$\lambda\lambda$1661,1666, respectively, for simplicity.
Thirty-five and 7 of our sample galaxies have $S/N>3$ detections of [\neiii] and \ciii], respectively.
We find that one of our sample galaxies, GLASS\_150008, has a detection of \niii] with $S/N=4.2$.
We also identify the detection of \niv] ($S/N=4.3$) from one of our sample galaxies, CEERS\_01019, which has also been reported by \citet{Larson2023arXiv}.

\tcra{Figure \ref{fig:spec} shows JWST/NIRCam images\footnote{\tcra{The imaging data of GLASS\_150008 are drawn from the UNCOVER \citep{Bezanson2022arXiv} website (\url{https://jwst-uncover.github.io/}), while the imaging data of CEERS\_01019 are taken from the CEERS survey \citep{Finkelstein2022arXiv} and reduced by \citet{Harikane2023a}.}} and spectra around \niv], \oiii], \niii], and \ciii] of GLASS\_150008 and CEERS\_01019.
The \oiii], \niii], and \ciii] of GLASS\_150008 and the \niv] of CEERS\_01019 have the $S/N$ ratios larger than 4.
We note that GLASS\_150008 has the ratio of each line flux in the \niii]$\lambda$1750 quintet to the total flux of the quintet in agreement within a 1$\sigma$ error level with that based on the electron density of $\lesssim10^{4}$ cm$^{-3}$ derived from the observed high \ciii]$\lambda$1907/\ciii]$\lambda$1909 ratio.}

\subsection{Emission-line Diagnostics} \label{subsec:diag}
\begin{figure*}[t]
    \centering
    \includegraphics[width=18.0cm]{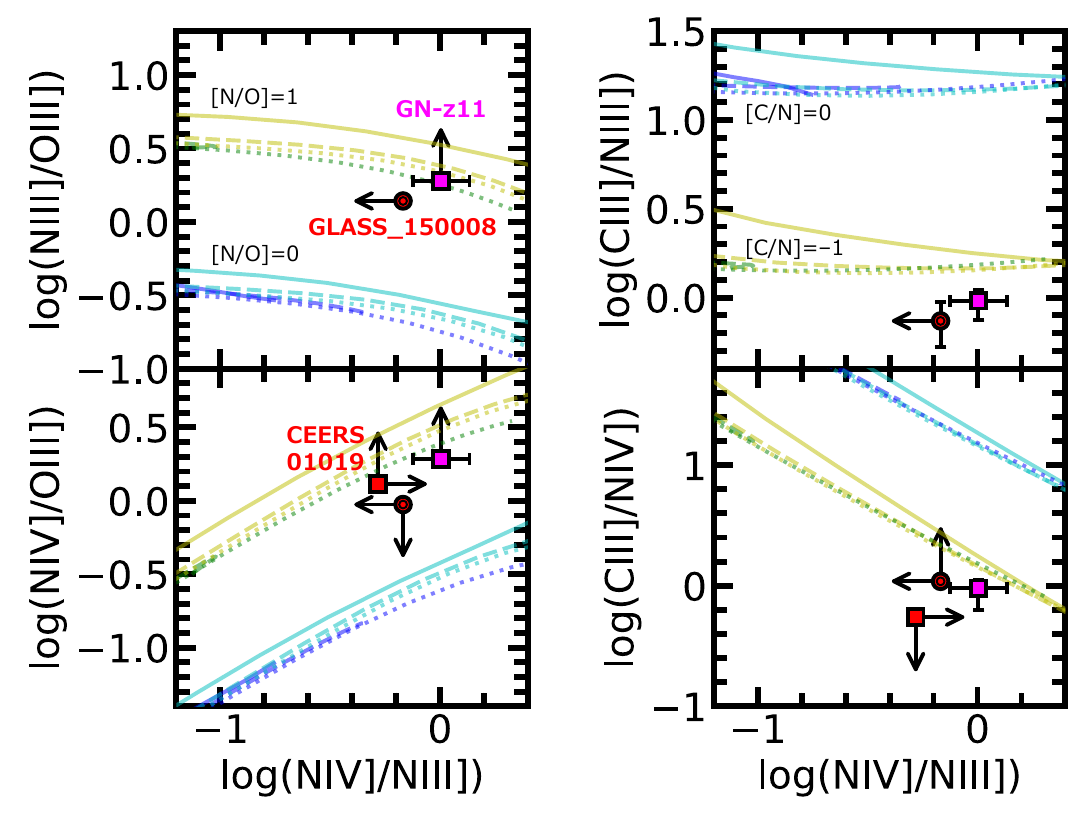}
    \caption{\niii]/\oiii] (top left), \niv]/\oiii] (bottom left), \ciii]/\niii] (top right), and \ciii]/\niv] (bottom right) ratios as a function of \niv]/\niii]. Observed emission-line ratios of GLASS\_150008, CEERS\_01019, and GN-z11 are represented by the double red circle, the red square, and the magenta square, respectively. The emission-line ratios of GN-z11 are measured by \citet{Maiolino2023arXiv} and \citet{Bunker2023arXiv}.
    \tcra{The blue and cyan curves show the Cloudy \citep{Ferland2013} photoionization models with ([N/O], [C/N$])=(0$, 0) based on the stellar and AGN radiations, respectively. The green and yellow curves denote the photoionization models with ([N/O], [C/N$])=(1$, $-1$) based on the stellar and AGN radiations, respectively. Line styles of these curves (solid, dashed, and dotted) correspond to the assumed metallicities ($\log(Z/Z_{\odot})=0.25$, $-0.5$, and $-1$, respectively).}
    Detailed measurements of the N/O and C/N ratios are made in Sections \ref{subsec:abun} and \ref{subsec:agn}.}
    \label{fig:n3o3}
\end{figure*}

The left panels of Figure \ref{fig:n3o3} compare \niii]/\oiii] and \niv]/\oiii] ratios of GLASS\_150008 (double red circle) and CEERS\_01019 (red square) with those of GN-z11 (magenta square) as a function of \niv]/\niii].
The \niii] and \niv] fluxes from \citet{Maiolino2023arXiv} and the \oiii] upper limit from \citet{Bunker2023arXiv}\footnote{Originally reported as $2\sigma$ upper limit in \citet{Bunker2023arXiv} but converted to $3\sigma$ upper limit for consistency with our flux measurements in Section \ref{subsec:flux}.} are scaled by \ciii] lines.
We find that GLASS\_150008 and CEERS\_01019 have high \niii]/\oiii] and \niv]/\oiii] ratios, respectively, both of which are comparable to those of GN-z11, which is reported to have a super-solar N/O ratio (Section \ref{sec:intro}).
The right panels of Figure \ref{fig:n3o3} also show that GLASS\_150008 and CEERS\_01019 have low \ciii]/\niii] and \ciii]/\niv] ratios, respectively, both of which are comparable to those of GN-z11.

To compare the observed emission-line ratios, we construct photoionization models using Cloudy \citep{Ferland2013} for both stellar and AGN radiations.
To model young star-forming galaxies, we use BPASS \citep{Stanway2018} binary stellar radiations under the assumptions of the instantaneous star-formation history with the stellar age of 10 Myr, upper star mass cut of 100 $M_{\odot}$ with the \citet{Salpeter1955} initial mass function (IMF), and the hydrogen density $n_{\rm H}$ of 300 cm$^{-3}$.
The $n_{\rm H}$ value is inferred from a typical value of the electron density ($n_{\rm e}$) of $\gtrsim300$ cm$^{-3}$ in the ISM of $z>4$ galaxies \citep{Isobe2023}.
We also fix He/H and metal-to-oxygen ratios to be the solar abundances.
We vary O/H and ionization parameter ($U$) values within the ranges of $-2\leq[$O/H$]\leq1$ and $-3.5\leq\log(U)\leq-0.5$ in 0.25 and 0.25 increments, respectively.
We set the stellar metallicity equal to the nebular metallicity defined by the O/H ratio.
We refer to this model as the young stellar model, hereafter.
We also construct the photoionization models for extremely young star-forming galaxies with very massive stars (extremely young massive stellar model, hereafter), assuming the stellar age of 1 Myr and upper star mass cut of 300 $M_{\odot}$.
The other parameters are the same as those of the young stellar model.
We also compute the stellar photoionization models with [N/O$]=1$, whose other parameters are the same as those of the young stellar model and extremely young massive stellar model.

We also construct AGN photoionization models whose incident radiation is parameterized by the following 4 parameters: the big-bump temperature $T_{\rm BB}$, the X-ray to UV ratio $\alpha_{\rm OX}$, the low-energy slope of the big-bump component $\alpha_{\rm UV}$, and the X-ray component slope $\alpha_{\rm X}$.
Typical AGNs have $\alpha_{\rm OX}\sim-1.4$ \citep{Zamorani1981} and $\alpha_{\rm UV}\sim-0.5$ \citep{Francis1993,Elvis1994}.
We refer to the AGN model with the parameter set of $T_{\rm BB}=1.5\times10^{5}$ K, $\alpha_{\rm OX}=-1.4$, $\alpha_{\rm UV}=-0.5$, and $\alpha_{\rm X}=-1.0$ as the default AGN model.
The flux density per frequency $f_{\nu}$ is expressed by:
\begin{equation}
    \label{equ:agn}
    f_{\nu}=\nu^{\alpha_{\rm UV}}\exp(-h\nu/kT_{\rm BB})\exp(-kT_{\rm IR}/h\nu)+a\nu^{\alpha_{\rm X}},
\end{equation}
where $a$ is a coefficient corresponding to $\alpha_{\rm OX}$ and $T_{\rm IR}$ is an infrared cutoff temperature at $kT_{\rm IR}=0.01$ Ryd.
The continuum above 100 keV is also assumed to cut off by $\nu^{-2}$.
Using the AGN radiation, we calculate emission line ratios under the assumption of $n_{\rm H}=300$ cm$^{-3}$.
We vary O/H, $U$, and N/O within the ranges of $-2\leq[$O/H$]\leq1$, $-3.5\leq\log(U)\leq-0.5$, $0\leq[$N/O$]\leq2$, and $-1\leq[$C/O$]\leq0$ in 0.25, 0.25, 1, and 1 increments, respectively.
We also fix He/H and metal-to-oxygen ratios other than N/O to be the solar abundances.
Note that different assumptions of $n_{\rm H}$ from 10 to $10^{6}$ cm$^{-3}$ can change key emission-line ratios of \niii]/\oiii], \ciii]/\oiii]. and \ciii]/\niii] by only $\lesssim0.1$ dex.

In Figure \ref{fig:n3o3}, we plot \tcra{the young stellar models with [N/O$]=0$ and [C/N$]=0$ in blue, the defalut AGN models with [N/O$]=0$ and [C/N$]=0$ in cyan, the young stellar models with [N/O$]=1$ and [C/N$]=-1$ in green, and the default AGN models with [N/O$]=1$ and [C/N$]=-1$ in yellow.}
We find that \tcra{the high \niii]/\oiii] ratios of GLASS\_150008 and GN-z11, as well as the high \niv]/\oiii] ratios of CEERS\_01019 and GN-z11, are larger than those predicted by both the young stellar model and the default AGN model with [N/O$]=0$ (blue and cyan).}
Similarly, the low \ciii]/\niii] and \ciii]/\niv] ratios in the 3 galaxies are comparable to those of the models with [C/N$]=-1$ \tcra{(green and yellow)}.
These results indicate that the 3 galaxies have very high N/O and low C/N ratios significantly above and below the solar abundance, respectively.
These results are not likely affected by the metallicity changes from $\log(\rm Z/Z_{\odot})=-1$ (blue) to 0.25 (red) as shown in Figure \ref{fig:n3o3}, while for accuracy, we derive metallicities first in Sections \ref{subsec:prop} and \ref{subsec:agn}.

\subsection{Nebular Property} \label{subsec:prop}
We derive color excesses $E(B-V)$ from Balmer line ratios of H$\beta$/H$\alpha$, H$\gamma$/H$\alpha$, and H$\gamma$/H$\beta$ as many as possible by assuming the dust attenuation curve of \citet{Calzetti2000} and the case B recombination.
Intrinsic values of the Balmer line ratios are calculated by PyNeb (\citealt{Luridiana2015}; v1.1.15).
If none of the above Balmer line ratios are available, we use a median value of $E(B-V)=0.07$ for the other galaxies in our sample.
Using the $E(B-V)$ values and \citet{Calzetti2000}'s curve, we correct emission-line ratios for dust attenuation.

For the 9 galaxies with the measurements of electron temperatures of \oiii\ ($T_{\rm e}(\rm \oiii)$) by \citet{Nakajima2023}, we use the $T_{\rm e}$ values and their errors.
In addition, we derive $T_{\rm e}$(\oiii) values of 5 galaxies with [\oiii]$\lambda$4363 and/or \oiii]$\lambda$1666.
The other galaxies in our sample are assumed to have $T_{\rm e}$(\oiii$)=15000$ K with 1$\sigma$ uncertainty of 5000 K.
We calculate $T_{\rm e}$(\oii) from $T_{\rm e}$(\oiii) and the empirical relation of \citet{Garnett1992}.
We assume $n_{\rm e}$ values to be 300 cm$^{-3}$.

We derive O$^{+}$/H$^{+}$ from [\oii]/H$\beta$ and $T_{\rm e}$(\oii) and O$^{2+}$/H$^{+}$ from [\oiii]$\lambda\lambda$4959,5007/H$\beta$ and $T_{\rm e}$(\oiii).
In our sample galaxies where only [\oii] upper limits are present, we presume the smallest values of [\oii] fluxes to be the ones obtained from [\oiii] under the assumption of $\log(U)=-1$. For these galaxies, we also suppose the true and maximum values of [\oii] fluxes to be their 1$\sigma$ and 3$\sigma$ upper limits, respectively. For our sample galaxies where [\oii] falls outside the wavelength coverage of NIRSpec, we assume the true [\oii] fluxes of the galaxies to be [\oiii] divided by a median [\oiii]/[\oii] ratio of the other galaxies in our sample. The minimum and maximum values are assumed to be [\oiii] divided by [\oiii]/[\oii] at $\log(U)=-1$ and $-3$, respectively.
We then obtain gas-phase metallicity $12+\log(\rm O/H)$ from the equation ${\rm O/H}={\rm O}^{+}/{\rm H}^{+}+{\rm O}^{2+}/{\rm H}^{+}$ by ignoring neutral oxygen and O$^{3+}$ and higher-order oxygen ions as in e.g., \citet{Izotov2006}.

We include errors of the used fluxes as well as uncertainties of the assumptions of $T_{\rm e}$ and [\oii] for some of our sample galaxies into $12+\log(\rm O/H)$ errors by Monte Carlo simulations.
We derive 1000 values of $12+\log(\rm O/H)$ with $T_{\rm e}$ and flux values such as [\oii] randomly fluctuated by their errors under the assumption of the normal distribution.
Then, we derive the 16th and 84th percentiles of the distribution of the 1000 $12+\log(\rm O/H)$ values as the $\pm1\sigma$ confidence interval of $12+\log(\rm O/H)$.
We note that we cannot measure $12+\log(\rm O/H)$ of 7 galaxies in our sample due to the lack of H$\beta$.
The derived $T_{\rm e}$, $E(B-V)$, and $12+\log(\rm O/H)$ values are summarized in Table \ref{tab:abun}.
\tcra{We have confirmed that our derived $12+\log(\rm O/H)$ values of GLASS\_150008 and CEERS\_01019 are consistent with those measured by \citet{Jones2023arXiv} and \citet{Larson2023arXiv} based on the stellar radiation within a $1\sigma$ error level, respectively.}

\tcra{We also estimate $12+\log(\rm O/H)$ of GN-z11 in a similar way to our sample galaxies.
However, we use [\oiii]$\lambda$4363 instead of [\oiii]$\lambda$5007, which is outside the NIRSpec wavelength coverage in the case of GN-z11.
We assume $T_{\rm e}(\mathrm{\oiii})=15000\pm5000$ K and $n_{\rm e}=300$ cm$^{-3}$.
We estimate $E(B-V)=0.00$, $\log(U)=-2.04$, O$^{+}$/H$^{+}=6.08\times10^{-6}$, and O$^{2+}$/H$^{+}=9.39\times10^{-5}$ from the observed H$\delta$/H$\gamma$, [\oiii]$\lambda4363$/[\oii], [\oii]/H$\delta$, and [\oiii]$\lambda4363$/H$\delta$ ratios, respectively.
The H$\delta$/H$\gamma$ ratio is taken from \citet{Maiolino2023arXiv}, while the other line ratios are taken from \citet{Bunker2023arXiv}.
We obtain $12+\log(\rm O/H)=8.00^{+0.76}_{-0.46}$, which is consistent with that derived by \citet{Cameron2023btmp} and \citet{Senchyna2023arXiv} based on the stellar radiation within a $1\sigma$ error level.}

\subsection{Abundance Ratio \tcra{Based on Stellar Radiation}} \label{subsec:abun}
\begin{table*}[t]
    \begin{center}
    \caption{Atomic data}
    \label{tab:atom}
    \begin{tabular}{cccc} \hline \hline
		Ion & Emission process & Transition probability & Collision Strength \\
		(1) & (2) & (3) & (4) \\ \hline
		H$^{0}$ & Re & \citet{Storey1995} & --- \\
		C$^{2+}$ & CE & \citet{Wiese1996} & \citet{Berrington1985} \\
		N$^{2+}$ & CE & \citet{Galavis1998} & \citet{Blum1992} \\
		O$^{+}$ & CE & \citet{FroeseFischer2004} & \citet{Kisielius2009} \\
		O$^{2+}$ & CE & \citet{FroeseFischer2004} & \citet{Storey2014}$^{\dagger}$ \\
		Ne$^{2+}$ & CE & \citet{FroeseFischer2004} & \citet{McLaughlin2000} \\
		Ar$^{2+}$ & CE & \citet{MunozBurgos2009} & \citet{MunozBurgos2009} \\
		S$^{+}$ & CE & \citet{Rynkun2019} & \citet{Tayal2010} \\ \hline
    \end{tabular}
    \end{center}
    \tablecomments{(1): Ion. (2): Emission process. Re: Recombination; CE: Collisional excitation. (3): Reference of the transition probability. (4): Reference of the collision strength. $^{\dagger}$: \citet{Aggarwal1999} for \oiii].}
\end{table*}

\begin{table*}[t]
    \centering
    \caption{Abundance Ratio of GLASS\_150008, CEERS\_01019, and GN-z11}
    \begin{tabular}{ccccccc} \hline\hline
        ID & $12+\log(\rm O/H)$ & $\log(\rm N/O)$ & $\log(\rm C/O)$ & $\log(\rm C/N)$ & $\log(\rm Ne/O)$ & Rad. \\
        (1) & (2) & (3) & (4) & (5) & (6) & (7) \\ \hline
        GLASS\_150008 & $7.65^{+0.14}_{-0.08}$ & $-0.40^{+0.05}_{-0.07}$ & $-1.08^{+0.06}_{-0.14}$ & $-0.68^{+0.09}_{-0.15}$ & $\cdots$ & stellar \\
        \multirow{2}{*}{CEERS\_01019} & \tcra{$7.94^{+0.46}_{-0.31}$} & \tcra{$>0.28$} & \tcra{$<-1.04$} & \tcra{$<-1.32$} & \tcra{$-0.77^{+0.11}_{-0.10}$} & \tcra{stellar} \\
        & 8.23--8.50 & $>-0.01$ & $<-0.36$ & $<-0.35$ & $(-0.72)$--$(-0.48)$ & AGN \\
        \multirow{2}{*}{GN-z11} & \tcra{$8.00^{+0.76}_{-0.46}$} & \tcra{$>-0.36$} & \tcra{$>-1.01$} & \tcra{$-0.65^{+0.06}_{-0.12}$} & \tcra{$>-0.88$} & \tcra{stellar} \\
        & 8.58--9.23 & $>-0.09$ & $>-0.78$ & $(-0.80)$--$(-0.54)$ & $>-1.85$ & AGN \\ \hline
    \end{tabular}
    \tablecomments{(1) Name. (2) $12+\log(\rm O/H)$. (3)--(6) Abundance ratios. (7) Incident radiation assumed to derive \tcra{the abundance ratios}.}
    \label{tab:hlt}
\end{table*}

We derive Ne/O, C/O, and N/O ratios from ionic abundance ratios with similar ionization energies of Ne$^{2+}$/O$^{2+}$, C$^{2+}$/O$^{2+}$, and N$^{2+}$/O$^{2+}$ to minimize systematics of ionization correction factors (ICFs).
We calculate Ne$^{2+}$/O$^{2+}$ ratios from [\neiii]/[\oiii] ratios.
For GLASS\_150008 \tcra{and GN-z11} with the \oiii] detection, we use \ciii]/\oiii] and \niii]/\oiii] ratios to obtain C$^{2+}$/O$^{2+}$ and N$^{2+}$/O$^{2+}$ ratios with less systematics.
For the other galaxies in our sample, we calculate C$^{2+}$/O$^{2+}$ and N$^{2+}$/O$^{2+}$ ratios from [\oiii] lines.
We calculate these ionic abundance ratios using PyNeb with the same transition probabilities and collision strengths listed in Table \ref{tab:atom}.

To derive chemical abundance ratios (Ne/O, C/O, and N/O) from the ionic abundance ratios (Ne$^{2+}$/O$^{2+}$, C$^{2+}$/O$^{2+}$, and N$^{2+}$/O$^{2+}$), we extract ICFs from the young stellar model (Section \ref{subsec:diag}).
We have checked that our ICFs provide Ne/O and C/O ratios consistent with those based on \citet{Izotov2006} and \citet{Berg2019}'s ICFs, respectively.
We have also confirmed that ICF(Ne$^{2+}$/O$^{2+}$), ICF(C$^{2+}$/O$^{2+}$), and ICF(N$^{2+}$/O$^{2+}$) of all our sample galaxies are $\sim1$.
In addition, the extremely young massive stellar model provides Ne/O, C/O, and N/O values similar to those of the young stellar model.
We derive the errors of the Ne/O, C/O, and N/O ratios by Monte Carlo simulations in the same way as for $12+\log(\rm O/H)$ (Section \ref{subsec:prop}).
For our sample galaxies without the detections of [\neiii], \ciii], or \niii], we use $3\sigma$ upper limits of these lines to obtain upper limits of Ne/O, C/O, or N/O ratios.
\tcra{We list the derived abundance ratios in Table \ref{tab:abun}, while those of GLASS\_150008, CEERS\_01019, and GN-z11 are highlighted in Table \ref{tab:hlt}.}
\tcra{All the 3 galaxies have super-solar N/O values under the assumption of the stellar radiation.}


\subsection{Abundance Ratio Based on AGN Radiation} \label{subsec:agn}
\begin{figure*}[t]
    \centering
    \includegraphics[width=18.0cm]{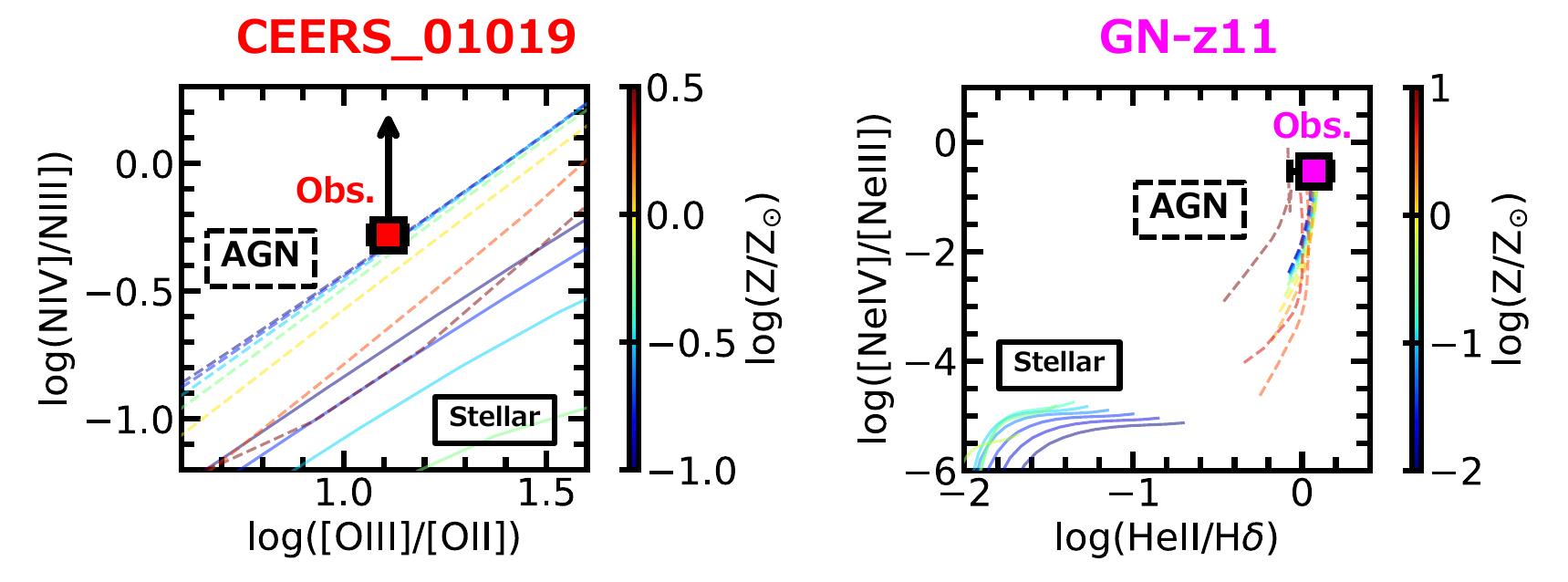}
    \caption{\tcra{(Left) \niv]/\niii] vs. [\oiii]/[\oii] of CEERS\_01019 (red square). (Right) [\neiv]/[\neiii] vs. \heii/H$\delta$ of GN-z11 (magenta square).} Cloudy photoionization models of young massive stellar population and AGN are shown by the solid and dashed lines, respectively, color-coded by metallicity. The observed emission-line ratios are not reproduced by the stellar models but the AGN models.}
    \label{fig:agn_stel}
\end{figure*}

\begin{figure}[t]
    \centering
    \includegraphics[width=8.0cm]{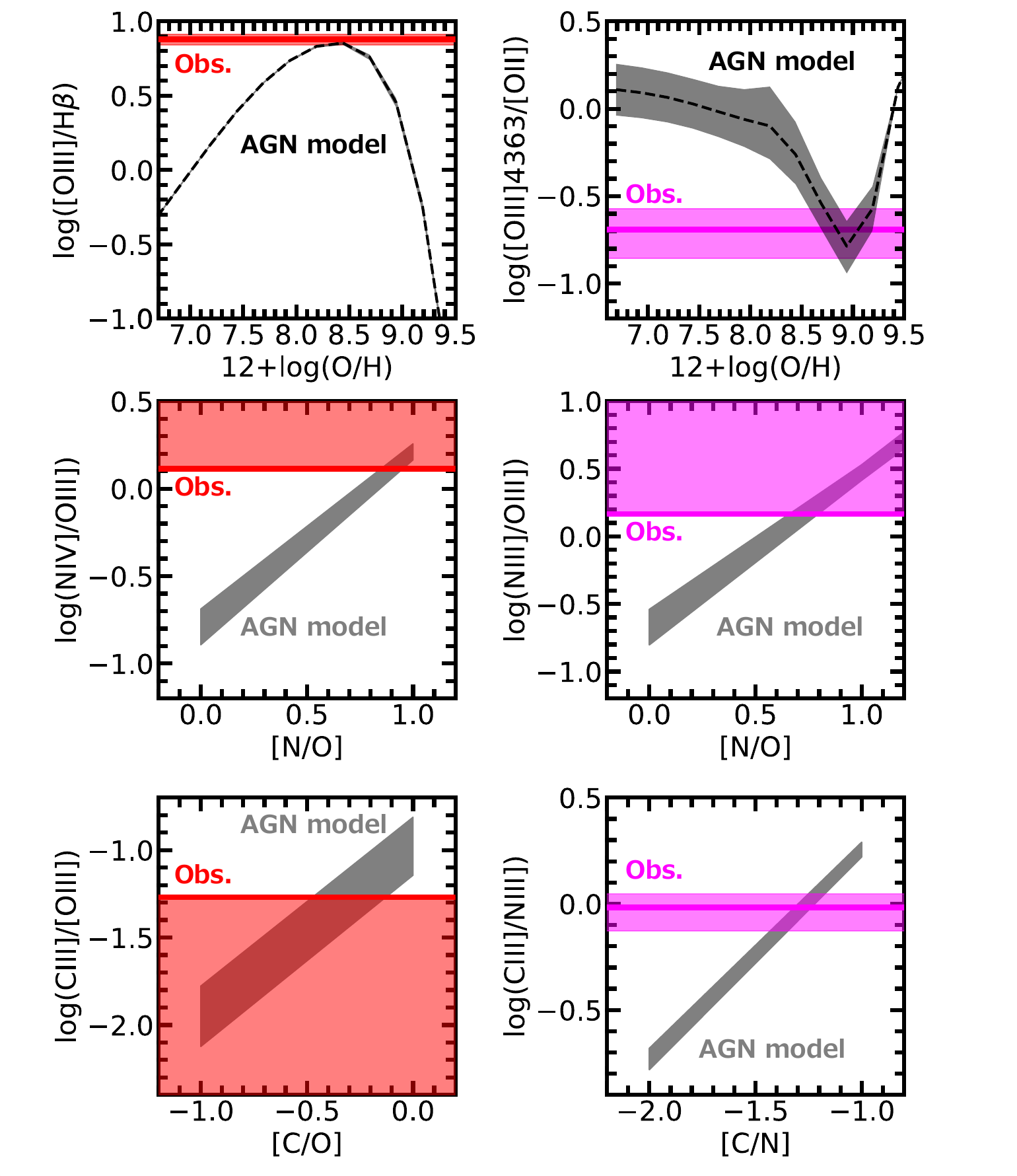}
    \caption{Comparison with the observed line ratios and the AGN photoionization models with different $12+\log(\rm O/H)$ (\tcra{top}), N/O (\tcra{middle}), C/O (bottom left), and C/N (bottom right). \tcra{The observed emission-line ratios of CEERS\_01019 and GN-z11 are shown by the red and magenta shaded regions, respectively. The black dashed lines in the top panels denote the emission-line ratios predicted by the AGN models with the $U$ values} determined by [\oiii]/[\oii] and \niv]/\niii] for CEERS\_01019 and GN-z11, respectively. The gray shaded regions \tcra{(in the top panels)} represent uncertainties of the observed ionization parameters, \tcra{while the gray shaded regions in the middle and bottom panels correspond to the uncertainties including both the $U$ and $12+\log(\rm O/H)$ uncertainties.}}
    \label{fig:agn}
\end{figure}

\begin{figure}[t]
    \centering
    \includegraphics[width=8.0cm]{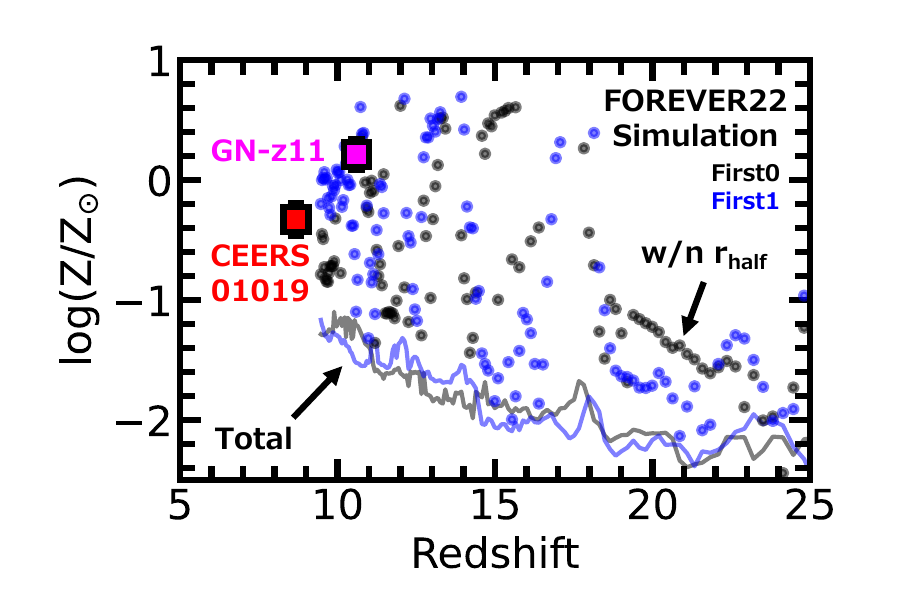}
    \caption{Metallicity as a function of redshift. The metallicities of CEERS\_01019 (red square) and GN-z11 (magenta square) are based on their O/H ratios. The black and blue symbols are FOREVER22 \citep{Yajima2022barXiv,Yajima2022} simulated galaxies of First0 and First1, respectively, while the dots and the solid lines denote metallicities measured within the half-mass radii ($r_{\rm half}$) and the total gas of the galaxies, respectively.}
    \label{fig:metz}
\end{figure}

\subsubsection{\tcra{AGN Possibility}} \label{subsubsec:agnpos}
\tcra{For accuracy of the abundance ratio measurements, we discuss the AGN possiblity of GLASS\_150008, CEERS\_01019, and GN-z11.
First,} GLASS\_150008 does not have very high ionization lines that require the presence of AGN.
We have also checked that GLASS\_150008 does not have a clear broad component in the H$\beta$ line profile.
We thus conclude that GLASS\_150008 does not have AGN signatures at least in the current data set.

Contrary to GLASS\_150008, \citet{Larson2023arXiv} have reported that CEERS\_01019 with the \niv] detection has an H$\beta$ broad component with the full-width half maximum of $\sim1200$ km s$^{-1}$, which can be an evidence of AGN.
\citet{Maiolino2023arXiv} have also reported that GN-z11 has a detection of [\neiv]$\lambda\lambda$2422,2424 lines that require high-energy photons beyond 63.5 eV, which can also suggest that GN-z11 is an AGN.

Here, we check if emission line ratios of CEERS\_01019 and GN-z11 cannot be reproduced by \tcra{the BPASS stellar photoionization models}.
The solid curves \tcra{in Figure \ref{fig:agn_stel}} represent predicted emission line ratios of the extremely young massive stellar model.
The model contains very massive stars up to 300 $M_{\odot}$, which are expected to produce harder radiation than normal stellar populations.
However, we confirm that even the extremely young massive stellar model cannot reproduce the observed \niv]/\niii] and [\oiii]/[\oii] relation of CEERS\_01019 (red \tcra{square}) or the observed [\neiv]$\lambda\lambda$2422,2424/[\neiii] and \niv]/\niii] relation of GN-z11 (magenta square; \citealt{Maiolino2023arXiv}).
\tcra{We can say that it is difficult to reproduce the observed emission lines of CEERS\_01019 and GN-z11 with the BPASS models.}

\tcra{This does not necessarily mean that all stellar photoionization models fail to reproduce the observed emission lines of CEERS\_01019 or GN-z11.
For example, \citet{Marques-Chaves2023arXiv} claim that the observed high \niv]/\niii] ratio of CEERS\_01019 can be explained by the 3MdB stellar photoionization model \citep{Morisset2015}.
However, as shown by the color-coded dashed lines in Figure \ref{fig:agn_stel}, we find that CEERS\_01019 has the lower limit of \niv]/\niii] and the [\oiii]/[\oii] ratio simultaneously reproduced by the default AGN model.}
We also find that both [\neiv]/[\neiii] and \heii/H$\delta$ ratios of GN-z11 measured by \citet{Maiolino2023arXiv} are reproduced by the default AGN model except for $\alpha_{\rm OX}=-1.5$ (low-$\alpha_{\rm OX}$ AGN model, hereafter).
\tcra{These results indicate the presence of AGN models that consistently reproduce the observed high ionization lines.}

\subsubsection{\tcra{Abundance Ratio Re-estimation of CEERS\_01019}} \label{subsubsec:c1019}
\tcra{Then, we} re-estimate $12+\log(\rm O/H)$ and N/O of CEERS\_01019 by using these AGN models.
The middle left panel of Figure \ref{fig:agn} shows the observed [\oiii]/H$\beta$ ratio of CEERS\_01019 (red solid line; 1$\sigma$ uncertainty is shown by the red shaded region) and those predicted by the default AGN models with the $U$ value derived from the observed [\oiii]/[\oii] ratio (gray dashed line; 1$\sigma$ uncertainty based on the observed [\oiii]/[\oii] uncertainty is shown by the gray shaded region) as a function of $12+\log(\rm O/H)$.
Note that [\oiii]/H$\beta$ values slightly decrease with increasing N/O.
With the default AGN models with [N/O$]=1$, we find that the observed and model-predicted [\oiii]/H$\beta$ ratios agree with each other at $12+\log(\rm O/H)=8.23$--8.50.
We then derive the N/O ratio of CEERS\_01019 by comparing the observed lower limit of \niv]/\oiii] with those predicted by the default AGN models with the metallicity range of $12+\log(\rm O/H)=8.23$--8.50, $U$ based on the observed [\oiii]/[\oii] ratio, and different N/O ratios.
We find that the observed lower limit of \niv]/\oiii] favors the model with the super-solar value of $[\rm N/O]>0.85$ (i.e., $\log(\rm N/O)>-0.01$).

We derive C/O and Ne/O ratios of CEERS\_01019 from \ciii]/[\oiii] and [\neiii]/[\oiii] ratios in the same way as for the N/O ratio, using the default AGN models with $12+\log(\rm O/H)=8.23$--8.50 and [N/O$]=1$.
We obtain [C/O$]<-0.13$ and [Ne/O$]=(-0.09)$--0.15, and our derived C/O and N/O ratios provide a low value of [C/N$]<-0.98$ (i.e., $\log(\rm C/N)<-0.35$).
Note that we can obtain a more strict upper limit of [C/N$]<-1.8$ based on the \ciii]/\niv] ratio.
The observed ratios among \niii], \niv], and \ciii] lines are simultaneously reproduced by the default AGN model with $12+\log(\rm O/H)=8.23$--8.50, $[\rm N/O]>0.85$, and [C/N$]<-1.8$.


\subsubsection{\tcra{Abundance Ratio Re-estimation of GN-z11}} \label{subsubsec:gnz11}
We derive $12+\log(\rm O/H)$ and N/O of GN-z11 in the similar way to CEERS\_01019, while we use the low-$\alpha_{\rm OX}$ AGN model, \niv]/\niii] \citep{Maiolino2023arXiv} as an indicator of $U$, the [\oiii]$\lambda$4363/[\oii] \citep{Bunker2023arXiv} for the metallicity measurement, and the lower limit of \niii]/\oiii] \citep{Maiolino2023arXiv,Bunker2023arXiv} for the N/O measurement.
Using the H$\delta$/H$\gamma$ ratio of \citet{Maiolino2023arXiv}, we derive $E(B-V)=0.00$ under the assumption of $T_{\rm e}=15000$ K and \citet{Calzetti2000}'s dust attenuation curve.
We obtain $12+\log(\rm O/H)=8.63$--9.21, which is higher than those based on the stellar radiation and [\oiii]$\lambda$4363 of $12+\log(\rm O/H)=7.03$--8.60 \citep{Cameron2023btmp}, $12+\log(\rm O/H)=7.84^{+0.06}_{-0.05}$ \citep{Senchyna2023arXiv}\tcra{, and our derived value of $12+\log(\rm O/H)=8.00^{+0.76}_{-0.46}$.}
This is because the AGN radiation can enhance [\oiii]$\lambda$4363, which results in overestimation of $T_{\rm e}$.

We derive C/N, C/O, and Ne/O ratios of GN-z11 from \ciii]/\niii] \citep{Maiolino2023arXiv}, \ciii]/\oiii] \citep{Maiolino2023arXiv,Bunker2023arXiv}, and [\neiii]/\oiii] \citep{Maiolino2023arXiv,Bunker2023arXiv} ratios in the same way as for the N/O ratio, using the low-$\alpha_{\rm OX}$ AGN models with $12+\log(\rm O/H)=8.23$--8.50 and [N/O$]=1$.
We obtain a low value of [C/N$]=(-1.43)$--($-1.17$), [C/O$]>-0.53$, and [Ne/O$]>-1.22$ (i.e., $\log(\rm C/N)=(-0.80)$--$(-0.54)$, $\log(\rm C/O)>-0.78$, and $\log(\rm Ne/O)>-1.85$).
We have confirmed that the C/N value based on \ciii]/\niii] is consistent with that based on \ciii]/\niv] within the error level.
We summarize the metallicities and the abundance ratios of GLASS\_150008, CEERS\_01019, and GN-z11 in Table \ref{tab:hlt}.

\tcra{Here, we check whether our derived metallicities of GN-z11 and CEERS\_01019 are too high at $z=10.6$ and 8.68, respectively.}
Figure \ref{fig:metz} compares the metallicity of GN-z11 (magenta square) \tcra{and CEERS\_01019 (red square)} based on the AGN radiation with those of First0 (black) and First1 (blue) galaxies of the cosmological hydrodynamics zoom-in simulation FOREVER22 \citep{Yajima2022barXiv,Yajima2022}.
\tcra{First0 and First1 galaxies have the same initial gas particle masses, dark matter particle masses, and gravitational softening lengths, but slightly different final halo masses of $3.5\times10^{11}$ and $3.2\times10^{11}\ M_{\odot}$, respectively \citep{Yajima2022barXiv}.}
Although the metallicities of the simulated galaxies calculated for the total gas (solid) are lower than that of GN-z11, we find that the simulated galaxies can instantaneously have super-solar metallicities within their half-mass radii ($\sim100$ pc at $z=10.6$, where GN-z11 is located).
This suggests that it is not very strange even for high-$z$ galaxies such as GN-z11 \tcra{and CEERS\_01019} to have high metallicities around the solar value.

\section{Nitrogen, Carbon, and Oxygen Abundances} \label{sec:nco}
\subsection{Result} \label{subsec:nco_res}
\begin{figure*}[t]
    \centering
    \includegraphics[width=18.0cm]{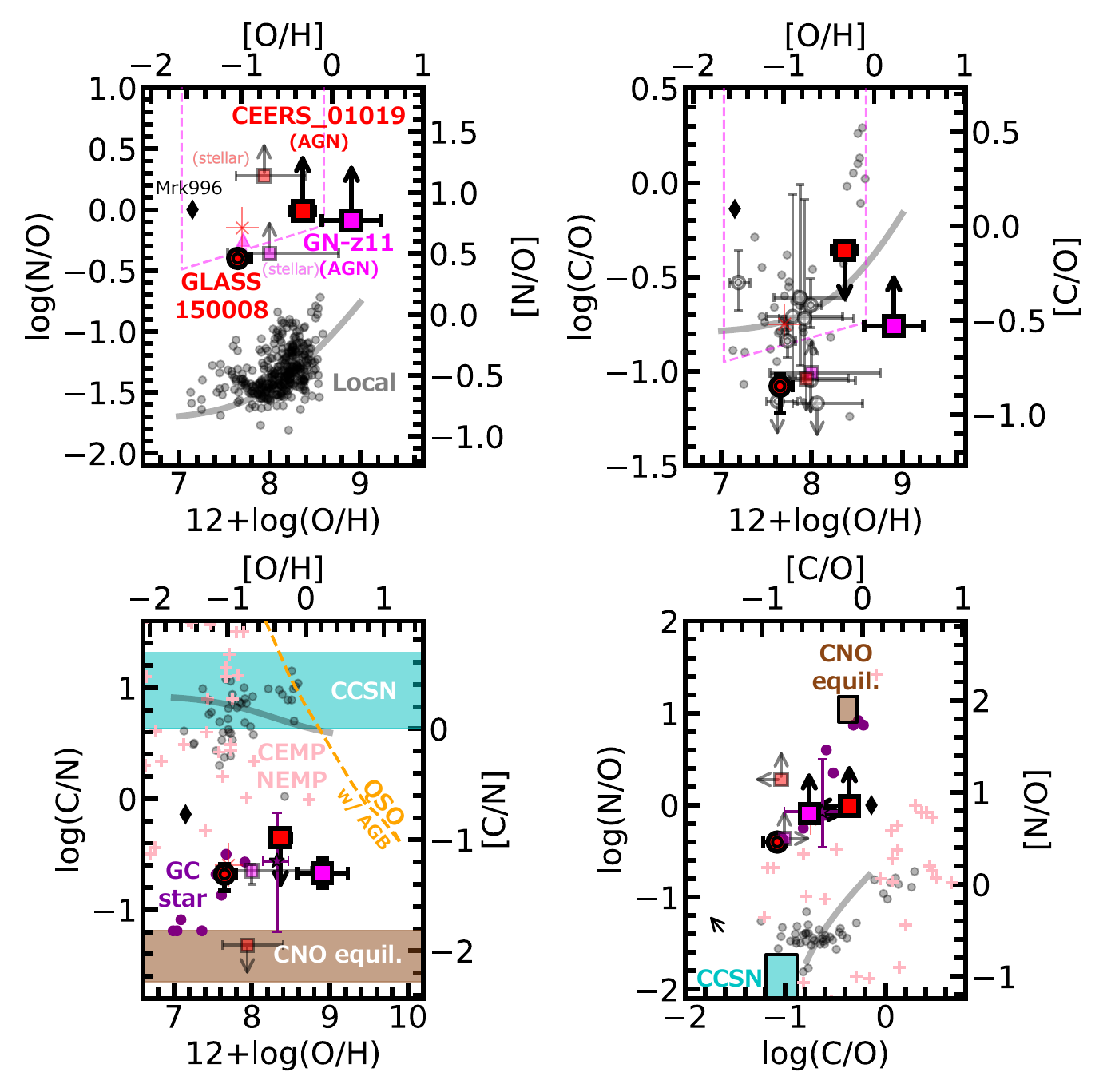}
    \caption{N/O (top left), C/O (top right), C/N (bottom left) as a function of $12+\log(\rm O/H)$. N/O as a function of C/O (bottom right). \tcra{The double red circle denotes the measurements of GLASS\_150008 based on the stellar photoionization model. The measurements of CEERS\_01019 (red square) and GN-z11 (magenta square) are based on the AGN photoionization models, while those based on the stellar photoionization models are shown by the semi-transparent red and magenta squares, respectively.
    The semi-transparent magenta dashed lines, magenta triangle, and red crosses are the abundance measurements of GN-z11 by \citet{Cameron2023btmp}, GN-z11 by \citet{Senchyna2023arXiv}, and CEERS\_01019 by \citet{Marques-Chaves2023arXiv}, respectively, all of which are based on the stellar radiation.} In the top right panel, C/O ratios of other galaxies in our sample are shown by \tcra{the semi-transparent white circles}. The double circles correspond to the measurements with $T_{\rm e}$ determinations. Mrk996, a Wolf-Rayet galaxy, is represented by the black diamonds \citep{Senchyna2023arXiv,Berg2016}. Local galaxies \citep{Berg2016,Berg2019,Izotov2006} and Galactic \hii\ regions \citep{Garcia-Rojas2007} are shown by the gray dots, while the gray curves are the empirical relation of the Galactic stars \citep{Nicholls2017}.
    The purple circles are dwarf stars in a globular cluster (GC) NGC6752 \citep{Carretta2005}, while those O/H values are taken from \citet{Senchyna2023arXiv}.
    \tcra{The purple stars show abundance ratio distributions of dwarf stars in a metal-rich GC 47 Tuc \citep{Briley2004,DOrazi2010}.}
    The pink \tcra{pluses} are carbon-enhanced metal-poor (CEMP) stars \citep{Norris2013} and nitrogen-enhanced metal-poor (NEMP) stars \citep{Beveridge1994}. The cyan and brown shaded regions indicate yields of CCSN \citep{Watanabe2023arXiv} and the equilibrium value of the CNO cycle \citep{Maeder2015}, respectively. The orange dashed line shows a chemical evolution model that reproduces emission-line ratios of quasars \citep{Hamman1993}, while nitrogen enrichment in the model is mainly caused by asymptotic giant branch (AGB) stars. The potential change of N/O and C/O by dust depletion \citep{Ferland2013} is indicated by the length of the small black arrow at the bottom left corner of the bottom right panel.}
    \label{fig:nco}
\end{figure*}

\begin{figure}[t]
    \centering
    \includegraphics[width=8.0cm]{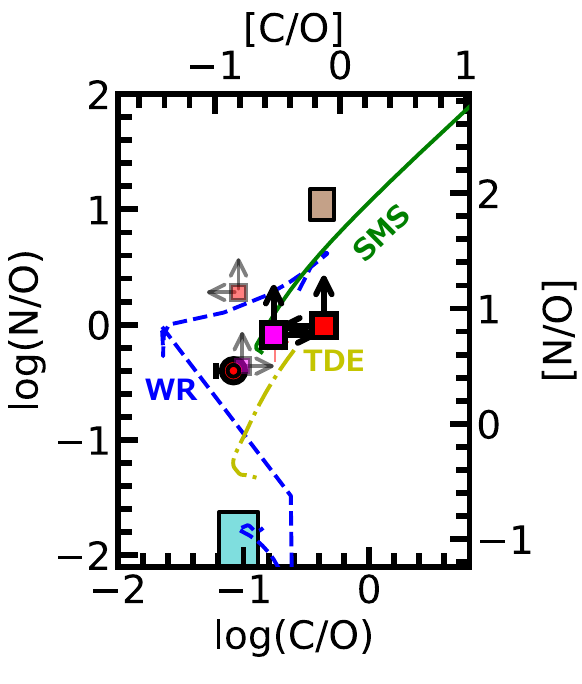}
    \caption{N/O as a function of C/O (bottom right). \tcra{The symbols are the same as in Figure \ref{fig:nco}, while} the blue dashed, green solid, and yellow dashed-dotted lines illustrate chemical evolution models of Wolf-Rayet (WR) stars, supermassive stars (SMS), and tidal disruption events (TDE), respectively, where most of massive stars are assumed to undergo the direct collapse (\citealt{Watanabe2023arXiv,Watanabeprep}).}
    \label{fig:nco_mod}
\end{figure}

\begin{figure*}[t]
    \centering
    \includegraphics[width=18.0cm]{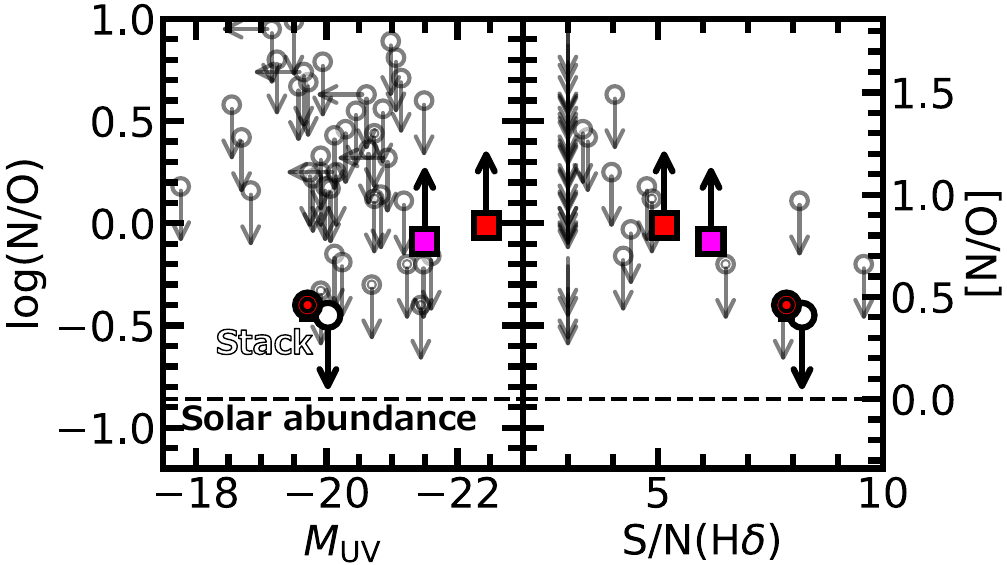}
    \caption{\tcra{N/O as a function of absolute UV magnitude $M_{\mathrm{UV}}$ (left) and $S/N$ ratio of H$\delta$ (right). The symbols are basically the same as those in the top right panel of Figure \ref{fig:nco}, but our sample galaxies with $S/N$(H$\delta)<3$ are shown by the gray arrows only in the right panel. The opaque white circle represents the N/O upper limit derived from the stacked spectrum of our sample galaxies. The black dashed line is the solar abundance.}}
    \label{fig:no_bri}
\end{figure*}

The top left panel of Figure \ref{fig:nco} shows N/O ratios as a function of metallicity.
We find that GLASS\_150008 and CEERS\_01019 show [N/O$]\gtrsim0.5$ as well as GN-z11.
\tcra{For CEERS\_01019 and GN-z11, this conclusion does not change regardless of whether AGN or stellar radiation is assumed.}
CEERS\_01019 and GLASS\_150008 are the second and third examples of super-solar N/O galaxies at $z>6$ after the report of GN-z11.
Hereafter we refer to these 3 galaxies (CEERS\_01019, GLASS\_150008, and GN-z11) as JWST N-rich galaxies.
The super-solar N/O values are significantly higher than those of typical local dwarf galaxies, Galactic \hii\ regions (gray dots), and typical Galactic stars (gray curve) at a given $12+\log(\rm O/H)$, but comparable to that of a Wolf-Rayet galaxy, Mrk996.

Contrary to the high N/O ratios, the top right panel of Figure \ref{fig:nco} illustrates that CEERS\_01019 \tcra{(together with many of our sample galaxies in the semi-transparent white circles)} has a C/O ratio comparable to those of the typical local galaxies, local stars, \tcra{and $z\sim3$ dwarf star-forming galaxies \citep[e.g.,][]{Llerena2022,Llerena2023}}.
GLASS\_150008 shows a C/O ratio even lower than those of the typical local galaxies and stars but comparable to some of our sample galaxies.

This discrepancy between the high N/O and low C/O ratios is illustrated by the bottom left panel of Figure \ref{fig:nco}, which shows the C/N ratios as a function of metallicity.
We find that all the JWST N-rich galaxies have [C/N] values less than $\sim-1$, which are significantly lower than those of the typical local galaxies and stars.
\tcra{This conclusion is also the case regardless of the assumed radiation.}

Such low C/N ratios are expected to be observed in nitrogen-loud quasars \citep[e.g.,][]{Batra2014}, while those O/H ratios have not fully been investigated.
Alternatively, we plot a chemical evolution model of \citet{Hamman1993} reproducing observed emission-line ratios of quasars.
Nitrogen in the model is mainly enriched by AGB stars.
Although the quasar model with AGB stars can produce low C/N ratios down to [C/N$]\sim-1$, it does not reproduce the (relatively-)low O/H ratios of the JWST N-rich galaxies simultaneously.
This is because AGB stars require much delay time ($\gtrsim1$ Gyr) to decrease the C/N ratio to [C/N$]\sim-1$, which increases O/H ratios too much.
This result also suggests that the JWST N-rich galaxies are not likely to be nitrogen-loud quasars whose nitrogen is enriched by AGB stars.

The bottom right panel of Figure \ref{fig:nco} shows the relations between N/O and C/O.
We confirm that the JWST N-rich galaxies have N/O ratios higher than those of the typical local galaxies and stars at a given C/O ratios, which indicate that only nitrogen is selectively enriched in the JWST N-rich galaxies with respect to the typical local galaxies and stars.

\subsection{Local Object with Similar C, N, O Abundances} \label{subsec:nco_loc}
We find that Mrk996, a local metal-poor dwarf star-forming galaxy with a high value of $\log(\rm N/O)\sim0$ \citep{Senchyna2023arXiv}, has a low C/N value similar to the JWST N-rich galaxies (bottom left panel of Figure \ref{fig:nco}).
The bottom right panel of Figure \ref{fig:nco} also shows that Mrk996 coincides with the JWST N-rich galaxies on the N/O vs. C/O plane.
Mrk996 is known to host WR stars evidenced by the presence of the WR bumps \citep[e.g.,][]{Telles2014}.
Similar to Mrk996, the JWST N-rich galaxies may also host WR stars.

We also identify some of carbon-enhanced metal-poor (CEMP) stars \citep{Norris2013} and nitrogen-enhanced metal-poor (NEMP) stars \citep{Beveridge1994} in the Galactic halo with the C/N vs. O/H and N/O vs. C/O relations comparable to those of the JWST N-rich galaxies as shown by the pink crosses in Figure \ref{fig:nco}.
It should be noted that some of such stars are giants undergoing the conversion of C to N in their hydrogen layers.

Moreover, the purple circles in the bottom panels of Figure \ref{fig:nco} indicate dwarf stars in a globular cluster (GC) NGC6752, which have been reported to have high N/O ratios similar to GN-z11 \citep{Senchyna2023arXiv}.
\tcra{The purple stars with the error bars also show median and 16th-84th percentile values of the abundance ratios of dwarf stars in a GC 47 Tuc.
Carbon and nitrogen abundances of the dwarf stars are drawn from \citet{Briley2004}, while the oxygen abundances are taken from \citet{DOrazi2010}.
The metallicity of GLASS\_150008 is similar to that of the dwarf stars in NGC6752, while CEERS\_01019 (GN-z11) has metallicity similar to (or close to) that of the dwarf stars in 47 Tuc.}
We find that the GC dwarf stars also have low C/N ratios and N/O vs. C/O relations similar to those of the JWST N-rich galaxies.
As surface abundance of dwarf stars are likely unevovled, the gas composition at the time of GC formation may also have similar abundance ratios to the JWST N-rich galaxies.
This implies that the JWST N-rich galaxies may be progenitors of GCs.

\subsection{Origin of Low C/N} \label{subsec:nco_org}
\begin{figure*}[t]
    \centering
    \includegraphics[width=18.0cm]{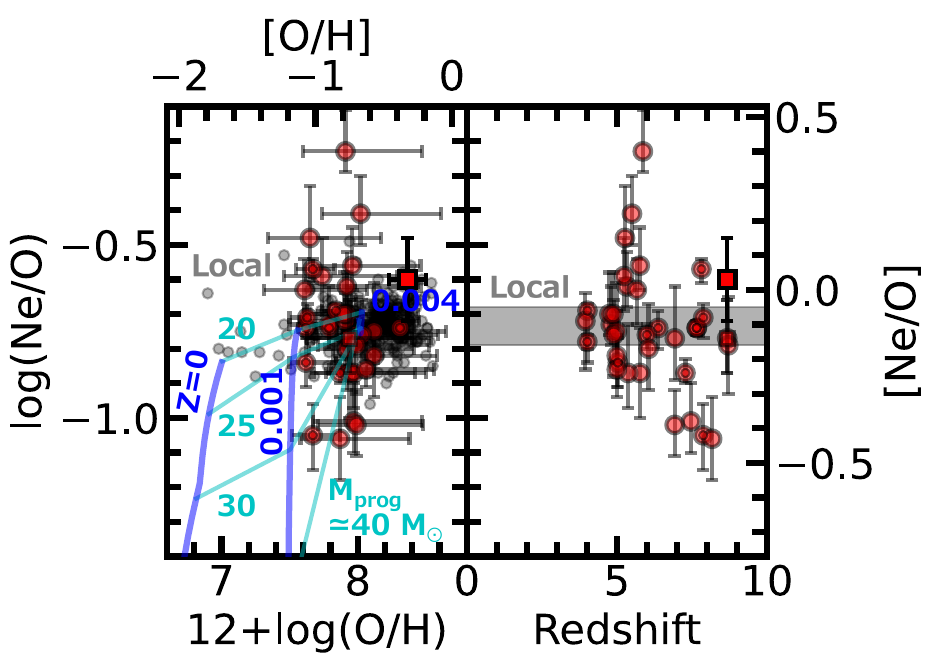}
    \caption{Ne/O ratio as a function of $12+\log(\rm O/H)$ (left) and redshift (right).
    \tcra{Our sample galaxies, except for CEERS\_01019 and GLASS\_150008, are the semi-transparent red circles. The double circles indicate the measurements with $T_{\rm e}$ determinations.
    CEERS\_01019 is shown by the opaque (AGN assumed) and semi-transparent (stellar assumed) red squares, while GLASS\_150008 does not have the Ne/O measurement (Table \ref{tab:hlt}).
    The gray dots in the left panel correspond to the local dwarf galaxies \citep{Izotov2006,Kojima2021,Isobe2022}, while the 16th and 84th percentiles of the Ne/O distribution of these local dwarf galaxies are shown by} the gray shaded region in the right panel. The blue curves represent chemical evolution models \citep{Watanabe2023arXiv} with different metallicities. The cyan lines denote isochrones of the chemical evolution models at the model ages of $10^{6.7}$, $10^{6.8}$, $10^{6.9}$, and $10^{7.0}$ year that correspond to the lifetime of progenitor stars with 40, 30, 25, and 20 $M_{\odot}$, respectively \citep{Portinari1998}.}
    \label{fig:neo}
\end{figure*}

In the bottom panels of Figure \ref{fig:nco}, we plot the abundance ratios processed by the CNO cycle at the equilibrium state \citep{Maeder2015} by the brown shaded regions, while core-collapse supernova yields \citep{Watanabe2023arXiv} are shown by the cyan shaded regions.
In contrast to the typical local galaxies and stars, we find that the JWST N-rich galaxies have the abundance ratios \tcrb{between those of CCSNe and} those of the equilibrium state of the CNO cycle.
\tcrb{This suggests that not only CCSNe but also the CNO cycle significantly contribute to the metal enrichment of the JWST N-rich galaxies.}
Thus, we need scenarios that can enrich gaseous materials originating from the CNO cycle while suppressing oxygen supply by CCSNe in the relatively low-metallicity environments.

In fact, these requirements are met by the scenarios presented to explain nitrogen enrichment of GN-z11 such as Wolf-Rayet (WR) stars \citep{Senchyna2023arXiv,Watanabe2023arXiv}, supermassive stars (SMS; \citealt{Charbonnel2023,Nagele2023}), and tidal disruption events (TDE; \citealt{Cameron2023btmp}).
In all the 3 scenarios, CNO-cycle material can be ejected from the outermost hydrogen-burning layer of stars via stellar winds (for the WR star and SMS scenarios) or gravitational interactions with black holes (for the TDE scenario).
Low-metallicity massive stars can also undergo direct collapse (i.e., no ejecta) due to the low opacity of the hydrogen envelope, resulting in a heavier helium core during collapse \citep{Heger2003}.
The blue dashed, green solid, and yellow dashed-dotted lines \tcra{in Figure \ref{fig:nco_mod}} represent chemical evolution models of WR stars, SMSs, and TDEs, respectively, all of which are constructed by \citet{Watanabe2023arXiv} and \citet{Watanabeprep}.
The WR star and SMS models include the yields of WR stars with 25--120 $M_{\odot}$ \citep{Limongi2018} and SMSs with $10^5\ M_{\odot}$ \citep{Nagele2023}, respectively.
The TDE model includes TDE yields calculated by \citet{Watanabeprep} with the nucleosynthesis code of \citet{Tominaga2007}, assuming the destruction of stars with 9--40 $M_{\odot}$.
Note that 100\%, 99.99\%, and 90\% of massive stars are assumed to undergo direct collapse (i.e., no ejecta) in the WR star, SMS, and TDE models, respectively, and that the remaining massive stars explode as CCSNe whose yields are taken from \citet{Nomoto2013}.
We confirm that all the 3 models can reproduce the observed N/O and C/O ratios of the JWST N-rich galaxies.
We have also checked that the N/O and C/O ratios of the wind yields from the $10^{3}\ M_{\odot}$ SMS \citep{Nagele2023} are also comparable to those of the JWST N-rich galaxies.
These findings suggest the presence of high-$z$ galaxies affected by WR stars, SMSs, and/or TDEs with frequent direct collapses.

\tcra{It should be noted that it is unclear that all high-$z$ galaxies follow these scenarios with \tcrb{the significant amount of} CNO-cycle materials because N/O ratios of the other high-$z$ galaxies are not constrained well.
Figure \ref{fig:no_bri} shows N/O ratios of our sample galaxies as a function of $M_{\mathrm{UV}}$ and $S/N$ ratio of H$\delta$.
The $M_{\mathrm{UV}}$ values of our sample galaxies and GN-z11 are taken from \citet{Nakajima2023} and \citet{Harikane2023carXiv}, respectively.
We find that all of our sample galaxies (semi-transparent white circle) but the JWST N-rich galaxies have poor upper limits above the solar abundance.
We also note that all the 3 JWST N-rich galaxies have large values of $M_{\mathrm{UV}}$ and/or $S/N$ ratio of H$\delta$ relative to the other galaxies in our sample.
These results suggest that we cannot tell whether the other galaxies in our sample have high N/O ratios consistent with the scenarios with dominant CNO-cycle materials, or whether the galaxies have low N/O ratios comparable to those of the CCSN yields, but are not constrained simply because the galaxies are not bright enough and/or the exposure time is not long enough.}

\tcra{We also investigate the typical N/O ratio of high-$z$ galaxies by stacking the JWST spectra.
We select JWST galaxy spectra with $R\sim1000$ and at $z=5.0$--9.4 to cover \oiii] and [\oiii].
The total number of such spectra reaches $\sim30$.
We perform median stacking on the $\sim30$ spectra and 100 times bootstrap resampling to estimate the error of the stacked spectrum.
We simply fit the Gaussian function to each emission line in the stacked spectrum.
We find that neither \niii] nor \oiii] is detected in the stacked spectrum.
We then estimate an upper limit of the N/O ratio from \niii]/[\oiii].
The (opaque) white circle in Figure \ref{fig:no_bri} shows the N/O upper limit, together with the median $M_{\mathrm{UV}}$ value of the used galaxies and the $S/N$ ratio of H$\delta$.
We find that the N/O upper limit of the stacked spectrum is still higher than the solar abundance value, which means that we cannot conclude whether even the typical N/O ratio of high-$z$ galaxies exceeds the solar abundance.}

\section{Neon Abundance} \label{sec:neo}
\subsection{Result} \label{subsec:neo_res}
The left panel of Figure \ref{fig:neo} shows Ne/O ratios of our sample galaxies (\tcra{semi-transparent red circle}) as a function of $12+\log(\rm O/H)$.
Although the majority of our sample galaxies have Ne/O ratios comparable to those of local dwarf galaxies (gray dot; \citealt{Izotov2006,Kojima2021,Isobe2022}), 4 of our sample galaxies (GLASS\_100003, CEERS\_00698, CEERS\_01143, and CEERS\_01149) have $\log(\rm Ne/O)$ values of $<-1.0$, which are significantly lower than those of the local dwarf galaxies and the other our sample galaxies.
The right panel of Figure \ref{fig:neo} illustrates the Ne/O ratios as a function of redshift.
Interestingly, all the 4 galaxies with the low Ne/O ratios are located at $z>6$.

\subsection{Origin of Low Ne/O} \label{subsec:neo_dis}
\tcra{Here, we explore the origins of the low Ne/O.
First, the CNO cycle cannot explain the low Ne/O ratios because the CNO cycle reduces the oxygen abundance and maintains the neon abundance.
The 4 galaxies with the low Ne/O ratios have only upper limits of N/O, which means that there is also a possibility that these 4 galaxies have low N/O ratios comparable to those of the CCSN yields (see also the last paragraph of Section \ref{subsec:nco_org}).
Since CCSNe can produce large amounts of neon and oxygen, we consider below the contribution of CCSNe to the low Ne/O ratios.}

We compare these data points with \citep{Watanabe2023arXiv}'s chemical evolution models accumulating the CCSN yields calculated by the nucleosynthesis code of \citet{Tominaga2007} under the assumption of \citet{Kroupa2001} IMF.
We assume the metallicity evolution of the Milky Way \citep{Suzuki2018}, and convert the maximum stellar ages of the models (Age$_{\rm mod}$) to $12+\log(\rm O/H)$.
The blue curves in Figure \ref{fig:neo} correspond to the chemical evolution models that accumulate CCSN ejecta with fixed metallicities of $Z=0$, 0.001, and 0.004.
We also illustrate isochrones of the models at $\log(\rm Age_{\rm mod}/yr)=6.7$, 6.8, 6.9, and 7.0, which correspond to lifetimes of stars with 40, 30, 25, and 20 $M_{\odot}$, respectively \citep{Portinari1998}.
As shown in the top left panel of Figure \ref{fig:neo}, the models predict an increase in Ne/O with Age$_{\rm mod}$.
This increase can originate from the prediction that CCSNe with more massive progenitors have higher temperatures in the carbon-burning layer \citep[e.g.,][]{Woosley2005}, which reduces neon abundance.

As shown in the top left panel of Figure \ref{fig:neo}, we find that the models with \tcra{$M_{\rm prog}\gtrsim30\ M_{\odot}$} reproduce the low Ne/O ratios of the 4 galaxies.
This \tcra{finding} agrees with the fact that the 4 galaxies are located at higher redshifts beyond 6, because galaxies at higher redshifts are expected to be younger and/or have top-heavy IMFs due to the metal-poorer environment.

\section{Summary} \label{sec:sum}
We present N/O, C/O, C/N, and Ne/O ratios of 70 star-forming galaxies at $z\sim4$--10, observed by the JWST/NIRSpec ERO, GLASS, and CEERS programs.
We derive these abundance ratios from emission-line ratios of similar ionization energies, accounting for both stellar and AGN radiation, particularly for CEERS\_01019 and GN-z11, which exhibit AGN signatures of high-ionization lines.
Our findings are summarized below:
\begin{itemize}
\item Among the 70 galaxies, we have pinpointed 2 galaxies with unique characteristics: CEERS\_01019 at $z=8.68$ and GLASS\_150008 at $z=6.23$. We find the low C/N and high N/O ratios these galaxies, which are also found in GN-z11, \tcrb{largely biased towards the C/N and N/O ratios at} the equilibrium of the CNO cycle. This suggests that these 3 galaxies \tcrb{contain the significant amount of} metals processed by the CNO cycle. \tcra{This conclusion does not change regardless of whether stellar or AGN radiation is assumed.}
\item The C/O and N/O ratios of these 3 galaxies are reproduced by chemical evolution models \tcrb{with the significant amount of} CNO-cycle materials involving Wolf-Rayet stars, supermassive stars, and tidal disruption events. Interestingly, these scenarios would require frequent direct collapses.
\item On the C/N vs. O/H plane, these 3 galaxies do not align with Galactic \hii\ regions, typical Galactic stars, typical star-forming galaxies, or nitrogen-loud quasars whose nitrogen is assumed to originate from AGB stars. They do, however, coincide with GC dwarf stars. We may see the site of GC formation in these 3 galaxies.
\item \tcra{The other galaxies in the ERO, GLASS, and CEERS programs have poor upper limits of N/O above the solar abundance. At this moment, we cannot tell whether these galaxies have super-solar N/O ratios.}
\item We identify 4 galaxies with very low Ne/O, with $\log(\rm Ne/O)<-1.0$. This low ratio indicates the abundance of massive ($\gtrsim30\ M_{\odot}$) stars.
\end{itemize}
\par
\par
We thank T. Nagao, A.K. Inoue, S. Aoyama\tcra{, H. Algera, T. Jones, M. Llerena, D. Schaerer, J. Scholtz, and A.P. Vijayan} for having useful discussions.
We are grateful to staff of the James Webb Space Telescope Help Desk for letting us know useful information.
This work is based on observations made with the NASA/ESA/CSA James Webb Space Telescope.
The data were obtained from the Mikulski Archive for Space Telescopes at the Space Telescope Science Institute, which is operated by the Association of Universities for Research in Astronomy, Inc., under NASA contract NAS 5-03127 for JWST.
These observations are associated with programs 1125, 2736, 1324, 1345\tcra{, and 2561}.
The authors acknowledge the ERO, GLASS, CEERS\tcra{, and UNCOVER} teams led by Klaus M. Pontoppidan, Tommaso Treu, Steven L. Finkelstein\tcra{, and Ivo Labbe}, respectively, for developing their observing programs with a zero-exclusive-access period.
This work is based on observations taken by the CANDELS Multi-Cycle Treasury Program with the NASA/ESA HST, which is operated by the Association of Universities for Research in Astronomy, Inc., under NASA contract NAS5-26555.
This work was supported by the joint research program of the Institute for Cosmic Ray Research (ICRR), University of Tokyo. 
This publication is based upon work
supported by the World Premier International Research
Center Initiative (WPI Initiative), MEXT, Japan.
Y.I., M.O., K.N., and Y.H. are supported by JSPS KAKENHI Grant Nos. 21J20785, 20H00180/21H04467, 20K22373, and 21K13953, respectively.
H.Y. is supported by MEXT/JSPS KAKENHI (21H04489) and JST FOREST Program (JP-MJFR202Z).
This research was supported by a grant from the Hayakawa Satio Fund awarded by the Astronomical Society of Japan.
\software{astropy \citep{Astropy2013,Astropy2018,Astropy2022}, PyNeb \citep{Luridiana2015}, Cloudy \citep{Ferland2013}}

\appendix
\restartappendixnumbering

\section{Table of Metallicity and Abundance Ratios \tcra{Based on the Stellar Radiation}} \label{sec:tab}

\startlongtable
\begin{deluxetable*}{cccccccccc}
\tablecaption{Metallicity and Abundance Ratios of $z\sim4$--9 Galaxies \tcra{Based on the Stellar Radiation}\label{tab:abun}}
\tablehead{
\colhead{\scriptsize  ID} & \colhead{\scriptsize  Redshift} & \colhead{\scriptsize  $T_{\rm e}$} & \colhead{\scriptsize  $E(B-V)$} & \colhead{\scriptsize  $12+\log(\rm O/H)$} & \colhead{\scriptsize  $\log(\rm Ne/O)$} & \colhead{\scriptsize  $\log(\rm C/O)$} & \colhead{\scriptsize  $\log(\rm N/O)$} & \colhead{\scriptsize  $\log(\rm Ar/O)$} & \colhead{\scriptsize  $\log(\rm S/O)$} \\
\colhead{\scriptsize  } & \colhead{\scriptsize  } & \colhead{\scriptsize  $10^{4}$ K} & \colhead{\scriptsize  mag} & \colhead{\scriptsize  } & \colhead{\scriptsize  } & \colhead{\scriptsize  } & \colhead{\scriptsize  } & \colhead{\scriptsize  } & \colhead{\scriptsize  }\\
\colhead{\scriptsize  (1)} & \colhead{\scriptsize  (2)} & \colhead{\scriptsize  (3)} & \colhead{\scriptsize  (4)} & \colhead{\scriptsize  (5)} & \colhead{\scriptsize  (6)} & \colhead{\scriptsize  (7)} & \colhead{\scriptsize  (8)} & \colhead{\scriptsize  (9)} & \colhead{\scriptsize  (10)}
}
\startdata
\scriptsize {ERO\_04590} & \scriptsize {8.4956} & \scriptsize {$2.08\pm0.26^{\rm a}$} & \scriptsize {0.12} & \scriptsize {$7.19^{+0.13\,{\rm g}}_{-0.10}$} & \scriptsize {$<-0.22$} & \scriptsize {$-0.53^{+0.17}_{-0.15}$} & \scriptsize {$\cdots$} & \scriptsize {$\cdots$} & \scriptsize {$\cdots$} \\
\scriptsize {ERO\_05144} & \scriptsize {6.3784} & \scriptsize {$1.64\pm0.16^{\rm a}$} & \scriptsize {0.00} & \scriptsize {$7.79^{+0.11\,{\rm f}}_{-0.10}$} & \scriptsize {$-0.74^{+0.05}_{-0.06}$} & \scriptsize {$\cdots$} & \scriptsize {$\cdots$} & \scriptsize {$\cdots$} & \scriptsize {$<-0.87$} \\
\scriptsize {ERO\_06355} & \scriptsize {7.6651} & \scriptsize {$1.13\pm0.08^{\rm a}$} & \scriptsize {0.13} & \scriptsize {$8.31^{+0.10\,{\rm f}}_{-0.09}$} & \scriptsize {$-0.74^{+0.02}_{-0.02}$} & \scriptsize {$\cdots$} & \scriptsize {$\cdots$} & \scriptsize {$\cdots$} & \scriptsize {$\cdots$} \\
\scriptsize {ERO\_08140} & \scriptsize {5.2751} & \scriptsize {$1.5\pm0.5^{\rm d}$} & \scriptsize {0.05} & \scriptsize {$7.77^{+0.45\,{\rm f}}_{-0.31}$} & \scriptsize {$<-0.90$} & \scriptsize {$\cdots$} & \scriptsize {$\cdots$} & \scriptsize {$<-1.75$} & \scriptsize {$-1.64^{+0.13}_{-0.10}$} \\
\scriptsize {ERO\_10612} & \scriptsize {7.6601} & \scriptsize {$2.04\pm0.16^{\rm a}$} & \scriptsize {0.00} & \scriptsize {$7.62^{+0.08\,{\rm f}}_{-0.06}$} & \scriptsize {$-0.74^{+0.02}_{-0.02}$} & \scriptsize {$\cdots$} & \scriptsize {$\cdots$} & \scriptsize {$\cdots$} & \scriptsize {$\cdots$} \\
\scriptsize {GLASS\_10000} & \scriptsize {7.8809} & \scriptsize {$1.83^{+0.32\,{\rm b}}_{-0.32}$} & \scriptsize {0.00} & \scriptsize {$7.63^{+0.18\,{\rm g}}_{-0.13}$} & \scriptsize {$-0.71^{+0.04}_{-0.06}$} & \scriptsize {$\cdots$} & \scriptsize {$\cdots$} & \scriptsize {$\cdots$} & \scriptsize {$\cdots$} \\
\scriptsize {GLASS\_100001} & \scriptsize {7.8737} & \scriptsize {$1.5\pm0.5^{\rm d}$} & \scriptsize {0.00} & \scriptsize {$7.90^{+0.52\,{\rm f}}_{-0.30}$} & \scriptsize {$<-0.83$} & \scriptsize {$\cdots$} & \scriptsize {$<0.46$} & \scriptsize {$\cdots$} & \scriptsize {$\cdots$} \\
\scriptsize {GLASS\_100003} & \scriptsize {7.8773} & \scriptsize {$1.98\pm0.37^{\rm a}$} & \scriptsize {0.00} & \scriptsize {$7.67^{+0.20\,{\rm f}}_{-0.15}$} & \scriptsize {$-1.05^{+0.09}_{-0.10}$} & \scriptsize {$\cdots$} & \scriptsize {$<-0.30$} & \scriptsize {$\cdots$} & \scriptsize {$\cdots$} \\
\scriptsize {GLASS\_10021} & \scriptsize {7.2863} & \scriptsize {$1.66\pm0.17^{\rm a}$} & \scriptsize {\tcra{0.00}} & \scriptsize {$7.87^{+0.12\,{\rm f}}_{-0.10}$} & \scriptsize {$-0.87^{+0.04}_{-0.03}$} & \scriptsize {$\cdots$} & \scriptsize {$<-0.40$} & \scriptsize {$\cdots$} & \scriptsize {$\cdots$} \\
\scriptsize {GLASS\_110000} & \scriptsize {5.7631} & \scriptsize {$1.5\pm0.5^{\rm d}$} & \scriptsize {0.13} & \scriptsize {$7.96^{+0.50\,{\rm f}}_{-0.29}$} & \scriptsize {$-0.56^{+0.11}_{-0.07}$} & \scriptsize {$<-0.66$} & \scriptsize {$<0.25$} & \scriptsize {$<-1.92$} & \scriptsize {$<-1.57$} \\
\scriptsize {GLASS\_150008} & \scriptsize {6.2291} & \scriptsize {$1.93^{+0.12\,{\rm c}}_{-0.12}$} & \scriptsize {0.07$^{\rm e}$} & \scriptsize {$7.65^{+0.14\,{\rm h}}_{-0.08}$} & \scriptsize {$\cdots$} & \scriptsize {\tcra{$-1.08^{+0.06\,{\rm i}}_{-0.14}$}} & \scriptsize {\tcra{$-0.40^{+0.05\,{\rm j}}_{-0.07}$}} & \scriptsize {$\cdots$} & \scriptsize {$\cdots$} \\
\scriptsize {GLASS\_150029} & \scriptsize {4.5837} & \scriptsize {$1.76\pm0.14^{\rm a}$} & \scriptsize {0.16} & \scriptsize {$7.73^{+0.09\,{\rm f}}_{-0.08}$} & \scriptsize {$\cdots$} & \scriptsize {$-0.84^{+0.10}_{-0.09}$} & \scriptsize {$\cdots$} & \scriptsize {$<-2.63$} & \scriptsize {$-1.69^{+0.09}_{-0.11}$} \\
\scriptsize {GLASS\_160122} & \scriptsize {5.3312} & \scriptsize {$1.5\pm0.5^{\rm d}$} & \scriptsize {0.00} & \scriptsize {$7.92^{+0.54\,{\rm f}}_{-0.28}$} & \scriptsize {$-0.62^{+0.10}_{-0.06}$} & \scriptsize {$-0.72^{+0.63}_{-0.31}$} & \scriptsize {$<-0.03$} & \scriptsize {$<-1.96$} & \scriptsize {$<-1.26$} \\
\scriptsize {GLASS\_160133} & \scriptsize {4.0151} & \scriptsize {$1.48\pm0.07^{\rm a}$} & \scriptsize {0.13} & \scriptsize {$7.95^{+0.05\,{\rm f}}_{-0.05}$} & \scriptsize {$-0.78^{+0.02}_{-0.02}$} & \scriptsize {$\cdots$} & \scriptsize {$\cdots$} & \scriptsize {$\cdots$} & \scriptsize {$<-2.09$} \\
\scriptsize {GLASS\_40066} & \scriptsize {4.0197} & \scriptsize {$1.66^{+0.14\,{\rm b}}_{-0.14}$} & \scriptsize {0.23} & \scriptsize {$7.84^{+0.17\,{\rm h}}_{-0.10}$} & \scriptsize {$-0.69^{+0.02}_{-0.15}$} & \scriptsize {$\cdots$} & \scriptsize {$\cdots$} & \scriptsize {$<-2.76$} & \scriptsize {$-1.94^{+0.05}_{-0.60}$} \\
\scriptsize {GLASS\_50002} & \scriptsize {5.1333} & \scriptsize {$1.5\pm0.5^{\rm d}$} & \scriptsize {0.52} & \scriptsize {$8.32^{+0.51\,{\rm f}}_{-0.29}$} & \scriptsize {$<-0.37$} & \scriptsize {$\cdots$} & \scriptsize {$\cdots$} & \scriptsize {$<-1.95$} & \scriptsize {$<-1.78$} \\
\scriptsize {GLASS\_50038} & \scriptsize {5.7720} & \scriptsize {$1.5\pm0.5^{\rm d}$} & \scriptsize {0.00} & \scriptsize {$7.94^{+0.48\,{\rm f}}_{-0.29}$} & \scriptsize {$-0.87^{+0.13}_{-0.13}$} & \scriptsize {$<-1.00$} & \scriptsize {$<-0.15$} & \scriptsize {$\cdots$} & \scriptsize {$\cdots$} \\
\scriptsize {GLASS\_80029} & \scriptsize {3.9510} & \scriptsize {$1.5\pm0.5^{\rm d}$} & \scriptsize {0.09} & \scriptsize {$7.91^{+0.54\,{\rm h}}_{-0.33}$} & \scriptsize {$-0.72^{+0.08}_{-0.14}$} & \scriptsize {$\cdots$} & \scriptsize {$\cdots$} & \scriptsize {$-2.19^{+0.12}_{-0.38}$} & \scriptsize {$-1.62^{+0.08}_{-0.62}$} \\
\scriptsize {GLASS\_80070} & \scriptsize {4.7969} & \scriptsize {$1.5\pm0.5^{\rm d}$} & \scriptsize {0.11} & \scriptsize {$7.79^{+0.55\,{\rm f}}_{-0.29}$} & \scriptsize {$-0.73^{+0.11}_{-0.06}$} & \scriptsize {$-0.71^{+0.65}_{-0.32}$} & \scriptsize {$<0.18$} & \scriptsize {$<-2.26$} & \scriptsize {$<-1.80$} \\
\scriptsize {CEERS\_00323} & \scriptsize {5.6657} & \scriptsize {$1.5\pm0.5^{\rm d}$} & \scriptsize {0.07} & \scriptsize {$7.61^{+0.46\,{\rm g}}_{-0.30}$} & \scriptsize {$-0.63^{+0.09}_{-0.10}$} & \scriptsize {$<-0.42$} & \scriptsize {$<0.42$} & \scriptsize {$<-1.28$} & \scriptsize {$<-0.55$} \\
\scriptsize {CEERS\_00355} & \scriptsize {6.0998} & \scriptsize {$1.5\pm0.5^{\rm d}$} & \scriptsize {0.01} & \scriptsize {$7.88^{+0.59\,{\rm f}}_{-0.27}$} & \scriptsize {$<-0.94$} & \scriptsize {$<-0.72$} & \scriptsize {$<0.22$} & \scriptsize {$<-1.60$} & \scriptsize {$<-1.29$} \\
\scriptsize {CEERS\_00397} & \scriptsize {6.0005} & \scriptsize {$1.52^{+0.16\,{\rm b}}_{-0.17}$} & \scriptsize {0.10} & \scriptsize {$7.99^{+0.12\,{\rm f}}_{-0.10}$} & \scriptsize {$-0.76^{+0.03}_{-0.04}$} & \scriptsize {$-0.65^{+0.14}_{-0.12}$} & \scriptsize {$<-0.20$} & \scriptsize {$<-2.18$} & \scriptsize {$<-1.80$} \\
\scriptsize {CEERS\_00403} & \scriptsize {5.7609} & \scriptsize {$1.5\pm0.5^{\rm d}$} & \scriptsize {0.53} & \scriptsize {$8.12^{+0.46\,{\rm f}}_{-0.28}$} & \scriptsize {$<-0.62$} & \scriptsize {$<0.40$} & \scriptsize {$<1.46$} & \scriptsize {$<-2.03$} & \scriptsize {$-1.76^{+0.14}_{-0.14}$} \\
\scriptsize {CEERS\_00407} & \scriptsize {7.0291} & \scriptsize {$1.5\pm0.5^{\rm d}$} & \scriptsize {0.00} & \scriptsize {$7.83^{+0.52\,{\rm g}}_{-0.30}$} & \scriptsize {$<-0.42$} & \scriptsize {$<-0.27$} & \scriptsize {$<0.58$} & \scriptsize {$\cdots$} & \scriptsize {$\cdots$} \\
\scriptsize {CEERS\_00515} & \scriptsize {5.6644} & \scriptsize {$1.5\pm0.5^{\rm d}$} & \scriptsize {0.27} & \scriptsize {$7.68^{+0.47\,{\rm g}}_{-0.30}$} & \scriptsize {$<-0.35$} & \scriptsize {$<0.25$} & \scriptsize {$<1.26$} & \scriptsize {$<-1.21$} & \scriptsize {$<-0.72$} \\
\scriptsize {CEERS\_00603} & \scriptsize {6.0568} & \scriptsize {$1.5\pm0.5^{\rm d}$} & \scriptsize {0.04} & \scriptsize {$7.92^{+0.49\,{\rm f}}_{-0.28}$} & \scriptsize {$-0.80^{+0.13}_{-0.12}$} & \scriptsize {$<-0.65$} & \scriptsize {$<0.18$} & \scriptsize {$<-1.50$} & \scriptsize {$<-1.09$} \\
\scriptsize {CEERS\_00618} & \scriptsize {6.0499} & \scriptsize {$1.5\pm0.5^{\rm d}$} & \scriptsize {0.00} & \scriptsize {$7.55^{+0.51\,{\rm g}}_{-0.30}$} & \scriptsize {$<-0.31$} & \scriptsize {$<-0.14$} & \scriptsize {$<0.67$} & \scriptsize {$<-0.86$} & \scriptsize {$<-0.39$} \\
\scriptsize {CEERS\_00670} & \scriptsize {5.8045} & \scriptsize {$1.5\pm0.5^{\rm d}$} & \scriptsize {0.34} & \scriptsize {$7.87^{+0.55\,{\rm g}}_{-0.29}$} & \scriptsize {$<-0.39$} & \scriptsize {$<0.29$} & \scriptsize {$<1.31$} & \scriptsize {$<-1.29$} & \scriptsize {$\cdots$} \\
\scriptsize {CEERS\_00689} & \scriptsize {7.5458} & \scriptsize {$1.5\pm0.5^{\rm d}$} & \scriptsize {0.07$^{\rm e}$} & \scriptsize {$7.95^{+0.47\,{\rm f}}_{-0.30}$} & \scriptsize {$<-0.80$} & \scriptsize {$<-0.18$} & \scriptsize {$<0.32$} & \scriptsize {$\cdots$} & \scriptsize {$\cdots$} \\
\scriptsize {CEERS\_00698} & \scriptsize {7.4708} & \scriptsize {$1.5\pm0.5^{\rm d}$} & \scriptsize {0.00} & \scriptsize {$7.97^{+0.50\,{\rm f}}_{-0.30}$} & \scriptsize {$-1.01^{+0.11}_{-0.09}$} & \scriptsize {$<-0.91$} & \scriptsize {$<-0.16$} & \scriptsize {$\cdots$} & \scriptsize {$\cdots$} \\
\scriptsize {CEERS\_00707} & \scriptsize {4.8964} & \scriptsize {$1.5\pm0.5^{\rm d}$} & \scriptsize {0.24} & \scriptsize {$8.03^{+0.56\,{\rm h}}_{-0.34}$} & \scriptsize {$-0.76^{+0.10}_{-0.12}$} & \scriptsize {$<-0.82$} & \scriptsize {$<0.25$} & \scriptsize {$<-2.42$} & \scriptsize {$-2.00^{+0.09}_{-0.59}$} \\
\scriptsize {CEERS\_00717} & \scriptsize {6.9315} & \scriptsize {$1.5\pm0.5^{\rm d}$} & \scriptsize {0.21} & \scriptsize {$7.95^{+0.59\,{\rm f}}_{-0.28}$} & \scriptsize {$-0.77^{+0.15}_{-0.12}$} & \scriptsize {$<-0.31$} & \scriptsize {$<0.60$} & \scriptsize {$\cdots$} & \scriptsize {$\cdots$} \\
\scriptsize {CEERS\_00792} & \scriptsize {6.2572} & \scriptsize {$1.5\pm0.5^{\rm d}$} & \scriptsize {0.00} & \scriptsize {$8.10^{+0.58\,{\rm f}}_{-0.29}$} & \scriptsize {$<-0.87$} & \scriptsize {$<-0.67$} & \scriptsize {$<0.16$} & \scriptsize {$<-1.54$} & \scriptsize {$<-1.18$} \\
\scriptsize {CEERS\_01019} & \scriptsize {8.6791} & \scriptsize {$1.5\pm0.5^{\rm d}$} & \scriptsize {0.00} & \scriptsize {$7.94^{+0.46\,{\rm f}}_{-0.31}$} & \scriptsize {$-0.77^{+0.11}_{-0.10}$} & \scriptsize {$<-1.04$} & \scriptsize {\tcra{$>0.28^{\rm k}$}} & \scriptsize {$\cdots$} & \scriptsize {$\cdots$} \\
\scriptsize {CEERS\_01023} & \scriptsize {7.7764} & \scriptsize {$1.5\pm0.5^{\rm d}$} & \scriptsize {0.07$^{\rm e}$} & \scriptsize {$7.67^{+0.55\,{\rm g}}_{-0.27}$} & \scriptsize {$<-0.41$} & \scriptsize {$\cdots$} & \scriptsize {$<0.81$} & \scriptsize {$\cdots$} & \scriptsize {$\cdots$} \\
\scriptsize {CEERS\_01025} & \scriptsize {8.7140} & \scriptsize {$1.5\pm0.5^{\rm d}$} & \scriptsize {0.07$^{\rm e}$} & \scriptsize {$7.99^{+0.53\,{\rm f}}_{-0.28}$} & \scriptsize {$-0.79^{+0.14}_{-0.14}$} & \scriptsize {$\cdots$} & \scriptsize {$<0.63$} & \scriptsize {$\cdots$} & \scriptsize {$\cdots$} \\
\scriptsize {CEERS\_01027} & \scriptsize {7.8197} & \scriptsize {$1.79^{+0.28\,{\rm b}}_{-0.27}$} & \scriptsize {0.25} & \scriptsize {$7.67^{+0.16\,{\rm f}}_{-0.12}$} & \scriptsize {$-0.57^{+0.03}_{-0.04}$} & \scriptsize {$<-0.60$} & \scriptsize {$<0.12$} & \scriptsize {$\cdots$} & \scriptsize {$\cdots$} \\
\scriptsize {CEERS\_01029} & \scriptsize {8.6100} & \scriptsize {$1.5\pm0.5^{\rm d}$} & \scriptsize {0.07$^{\rm e}$} & \scriptsize {$7.92^{+0.54\,{\rm f}}_{-0.29}$} & \scriptsize {$<-0.64$} & \scriptsize {$<-0.16$} & \scriptsize {$<0.71$} & \scriptsize {$\cdots$} & \scriptsize {$\cdots$} \\
\scriptsize {CEERS\_01038} & \scriptsize {7.1945} & \scriptsize {$1.5\pm0.5^{\rm d}$} & \scriptsize {0.07$^{\rm e}$} & \scriptsize {$7.64^{+0.48\,{\rm g}}_{-0.28}$} & \scriptsize {$<-0.40$} & \scriptsize {$<0.00$} & \scriptsize {$<0.80$} & \scriptsize {$\cdots$} & \scriptsize {$\cdots$} \\
\scriptsize {CEERS\_01115} & \scriptsize {6.3002} & \scriptsize {$1.5\pm0.5^{\rm d}$} & \scriptsize {0.06} & \scriptsize {$7.70^{+0.57\,{\rm g}}_{-0.28}$} & \scriptsize {$<-0.66$} & \scriptsize {$<-0.47$} & \scriptsize {$<0.46$} & \scriptsize {$\cdots$} & \scriptsize {$<-0.34$} \\
\scriptsize {CEERS\_01143} & \scriptsize {6.9266} & \scriptsize {$1.5\pm0.5^{\rm d}$} & \scriptsize {0.00} & \scriptsize {$7.99^{+0.49\,{\rm f}}_{-0.28}$} & \scriptsize {$-1.02^{+0.10}_{-0.09}$} & \scriptsize {$<-1.05$} & \scriptsize {$<-0.19$} & \scriptsize {$\cdots$} & \scriptsize {$\cdots$} \\
\scriptsize {CEERS\_01149} & \scriptsize {8.1750} & \scriptsize {$1.5\pm0.5^{\rm d}$} & \scriptsize {0.02} & \scriptsize {$7.87^{+0.51\,{\rm g}}_{-0.29}$} & \scriptsize {$-1.06^{+0.12}_{-0.12}$} & \scriptsize {$-0.61^{+0.60}_{-0.36}$} & \scriptsize {$<0.14$} & \scriptsize {$\cdots$} & \scriptsize {$\cdots$} \\
\scriptsize {CEERS\_01160} & \scriptsize {6.5667} & \scriptsize {$1.5\pm0.5^{\rm d}$} & \scriptsize {0.00} & \scriptsize {$7.97^{+0.47\,{\rm f}}_{-0.28}$} & \scriptsize {$<-0.62$} & \scriptsize {$<-0.62$} & \scriptsize {$<0.25$} & \scriptsize {$\cdots$} & \scriptsize {$\cdots$} \\
\scriptsize {CEERS\_01173} & \scriptsize {4.9957} & \scriptsize {$1.5\pm0.5^{\rm d}$} & \scriptsize {0.31} & \scriptsize {$8.12^{+0.53\,{\rm f}}_{-0.27}$} & \scriptsize {$-0.75^{+0.14}_{-0.09}$} & \scriptsize {$<-0.27$} & \scriptsize {$<0.89$} & \scriptsize {$<-2.15$} & \scriptsize {$-1.83^{+0.12}_{-0.11}$} \\
\scriptsize {CEERS\_01236} & \scriptsize {4.4842} & \scriptsize {$1.5\pm0.5^{\rm d}$} & \scriptsize {0.00} & \scriptsize {$7.83^{+0.56\,{\rm f}}_{-0.27}$} & \scriptsize {$<-0.40$} & \scriptsize {$<-0.09$} & \scriptsize {$\cdots$} & \scriptsize {$<-1.43$} & \scriptsize {$<-1.33$} \\
\scriptsize {CEERS\_01289} & \scriptsize {4.8798} & \scriptsize {$1.5\pm0.5^{\rm d}$} & \scriptsize {0.07} & \scriptsize {$8.02^{+0.52\,{\rm f}}_{-0.28}$} & \scriptsize {$<-0.77$} & \scriptsize {$<-0.56$} & \scriptsize {$<0.43$} & \scriptsize {$<-1.81$} & \scriptsize {$<-1.33$} \\
\scriptsize {CEERS\_01294} & \scriptsize {4.9987} & \scriptsize {$1.5\pm0.5^{\rm d}$} & \scriptsize {0.15} & \scriptsize {$8.12^{+0.46\,{\rm f}}_{-0.30}$} & \scriptsize {$-0.82^{+0.13}_{-0.12}$} & \scriptsize {$<-0.45$} & \scriptsize {$<0.56$} & \scriptsize {$<-1.84$} & \scriptsize {$<-1.50$} \\
\scriptsize {CEERS\_01324} & \scriptsize {5.0074} & \scriptsize {$1.5\pm0.5^{\rm d}$} & \scriptsize {0.00} & \scriptsize {$7.68^{+0.54\,{\rm g}}_{-0.33}$} & \scriptsize {$<-0.29$} & \scriptsize {$<-0.19$} & \scriptsize {$<0.79$} & \scriptsize {$<-1.10$} & \scriptsize {$<-0.69$} \\
\scriptsize {CEERS\_01358} & \scriptsize {5.5038} & \scriptsize {$1.5\pm0.5^{\rm d}$} & \scriptsize {0.38} & \scriptsize {$8.02^{+0.59\,{\rm g}}_{-0.28}$} & \scriptsize {$-0.41^{+0.11}_{-0.09}$} & \scriptsize {$<0.31$} & \scriptsize {$<1.25$} & \scriptsize {$<-1.43$} & \scriptsize {$<-0.47$} \\
\scriptsize {CEERS\_01365} & \scriptsize {4.3074} & \scriptsize {$1.5\pm0.5^{\rm d}$} & \scriptsize {0.00} & \scriptsize {$7.71^{+0.49\,{\rm g}}_{-0.28}$} & \scriptsize {$<-0.49$} & \scriptsize {$\cdots$} & \scriptsize {$\cdots$} & \scriptsize {$<-1.26$} & \scriptsize {$<-0.76$} \\
\scriptsize {CEERS\_01374} & \scriptsize {4.9999} & \scriptsize {$1.5\pm0.5^{\rm d}$} & \scriptsize {0.13} & \scriptsize {$8.06^{+0.50\,{\rm f}}_{-0.27}$} & \scriptsize {$-0.86^{+0.10}_{-0.05}$} & \scriptsize {$<-1.17$} & \scriptsize {$<-0.20$} & \scriptsize {$\cdots$} & \scriptsize {$-1.84^{+0.08}_{-0.06}$} \\
\scriptsize {CEERS\_01401} & \scriptsize {5.3749} & \scriptsize {$1.5\pm0.5^{\rm d}$} & \scriptsize {0.07} & \scriptsize {$7.97^{+0.53\,{\rm f}}_{-0.29}$} & \scriptsize {$-0.87^{+0.14}_{-0.10}$} & \scriptsize {$<-0.76$} & \scriptsize {$<0.17$} & \scriptsize {$<-1.86$} & \scriptsize {$<-1.73$} \\
\scriptsize {CEERS\_01449} & \scriptsize {4.7519} & \scriptsize {$1.5\pm0.5^{\rm d}$} & \scriptsize {0.20} & \scriptsize {$7.92^{+0.52\,{\rm f}}_{-0.28}$} & \scriptsize {$-0.70^{+0.12}_{-0.06}$} & \scriptsize {$<-0.91$} & \scriptsize {$<0.11$} & \scriptsize {$-2.25^{+0.11}_{-0.25}$} & \scriptsize {$-1.78^{+0.07}_{-0.06}$} \\
\scriptsize {CEERS\_01465} & \scriptsize {5.2692} & \scriptsize {$1.5\pm0.5^{\rm d}$} & \scriptsize {0.00} & \scriptsize {$7.65^{+0.52\,{\rm f}}_{-0.31}$} & \scriptsize {$-0.48^{+0.15}_{-0.12}$} & \scriptsize {$<-0.22$} & \scriptsize {$<0.69$} & \scriptsize {$<-1.32$} & \scriptsize {$<-1.21$} \\
\scriptsize {CEERS\_01536} & \scriptsize {5.0337} & \scriptsize {$2.25\pm0.43^{\rm a}$} & \scriptsize {0.08} & \scriptsize {$7.62^{+0.22\,{\rm f}}_{-0.12}$} & \scriptsize {$-0.84^{+0.07}_{-0.07}$} & \scriptsize {$<-1.16$} & \scriptsize {$<-0.33$} & \scriptsize {$<-1.75$} & \scriptsize {$<-1.29$} \\
\scriptsize {CEERS\_01539} & \scriptsize {4.8841} & \scriptsize {$1.5\pm0.5^{\rm d}$} & \scriptsize {0.27} & \scriptsize {$7.90^{+0.50\,{\rm f}}_{-0.29}$} & \scriptsize {$-0.70^{+0.12}_{-0.07}$} & \scriptsize {$<-0.27$} & \scriptsize {$<0.74$} & \scriptsize {$<-2.18$} & \scriptsize {$-1.85^{+0.09}_{-0.07}$} \\
\scriptsize {CEERS\_01544} & \scriptsize {4.1901} & \scriptsize {$1.5\pm0.5^{\rm d}$} & \scriptsize {0.05} & \scriptsize {$7.76^{+0.49\,{\rm g}}_{-0.30}$} & \scriptsize {$<-0.31$} & \scriptsize {$<-0.01$} & \scriptsize {$\cdots$} & \scriptsize {$<-0.41$} & \scriptsize {$<-0.76$} \\
\scriptsize {CEERS\_01561} & \scriptsize {6.1967} & \scriptsize {$1.5\pm0.5^{\rm d}$} & \scriptsize {0.00} & \scriptsize {$7.65^{+0.52\,{\rm g}}_{-0.28}$} & \scriptsize {$<-0.69$} & \scriptsize {$<-0.51$} & \scriptsize {$<0.33$} & \scriptsize {$<-1.09$} & \scriptsize {$<-0.33$} \\
\scriptsize {CEERS\_01605} & \scriptsize {4.6309} & \scriptsize {$1.5\pm0.5^{\rm d}$} & \scriptsize {0.07} & \scriptsize {$7.86^{+0.48\,{\rm f}}_{-0.30}$} & \scriptsize {$<-0.75$} & \scriptsize {$\cdots$} & \scriptsize {$<0.55$} & \scriptsize {$<-1.79$} & \scriptsize {$<-1.71$} \\
\scriptsize {CEERS\_01658} & \scriptsize {4.6038} & \scriptsize {$1.5\pm0.5^{\rm d}$} & \scriptsize {0.24} & \scriptsize {$7.98^{+0.51\,{\rm f}}_{-0.30}$} & \scriptsize {$\cdots$} & \scriptsize {$<-0.42$} & \scriptsize {$\cdots$} & \scriptsize {$<-2.03$} & \scriptsize {$\cdots$} \\
\scriptsize {CEERS\_01677} & \scriptsize {5.8670} & \scriptsize {$1.5\pm0.5^{\rm d}$} & \scriptsize {0.00} & \scriptsize {$7.91^{+0.56\,{\rm f}}_{-0.31}$} & \scriptsize {$-0.23^{+0.12}_{-0.06}$} & \scriptsize {$<-0.21$} & \scriptsize {$<0.36$} & \scriptsize {$<-1.42$} & \scriptsize {$<-1.33$} \\
\scriptsize {CEERS\_02174} & \scriptsize {5.3030} & \scriptsize {$1.5\pm0.5^{\rm d}$} & \scriptsize {0.21} & \scriptsize {$8.08^{+0.53\,{\rm f}}_{-0.29}$} & \scriptsize {$<-0.39$} & \scriptsize {$\cdots$} & \scriptsize {$<0.95$} & \scriptsize {$<-1.60$} & \scriptsize {$<-1.54$} \\
\scriptsize {CEERS\_02782} & \scriptsize {5.2408} & \scriptsize {$1.5\pm0.5^{\rm d}$} & \scriptsize {0.32} & \scriptsize {$7.74^{+0.55\,{\rm g}}_{-0.28}$} & \scriptsize {$-0.59^{+0.11}_{-0.10}$} & \scriptsize {$<-0.01$} & \scriptsize {$<0.99$} & \scriptsize {$<-1.49$} & \scriptsize {$<-0.69$} \\
\enddata
\tablecomments{(1) ID. (2) Redshift. (3) Electron temperature. a: Based on [\oiii]$\lambda$4363 derived by \citet{Nakajima2023}; b: Based on [\oiii]$\lambda$4363 but not reported by \citet{Nakajima2023}; c: Based on \oiii]$\lambda$1666; d: Assumption. (4) Color excess. e: Assumed to be the median $E(B-V)$ of the other our sample galaxies. (5) Metallicity. f: [\oii] \tcra{$S/N\geq3$}; g: [\oii] $S/N<3$; h: [\oii] out of wavelength coverage. We include uncertainties of the assumptions of $T_{\rm e}$ and [\oii] into the errors of $12+\log(\rm O/H)$ and the abundance ratios by performing Monte Carlo simulations (Section \ref{subsec:abun}). \tcra{(6)--(10) Abundance ratios. i: Based on metallicity-insensitive \ciii]/\oiii]. The other C/O measurements are based on \ciii]/[\oiii]. j: Based on metallicity-insensitive \niii]/\oiii]. k: Based on \niv]/[\oiii] The other N/O measurements are based on \niii]/[\oiii].}}
\end{deluxetable*}

\section{Argon and Sulfur Abundances} \label{sec:ars}
\begin{figure}[h]
    \centering
    \includegraphics[width=8.0cm]{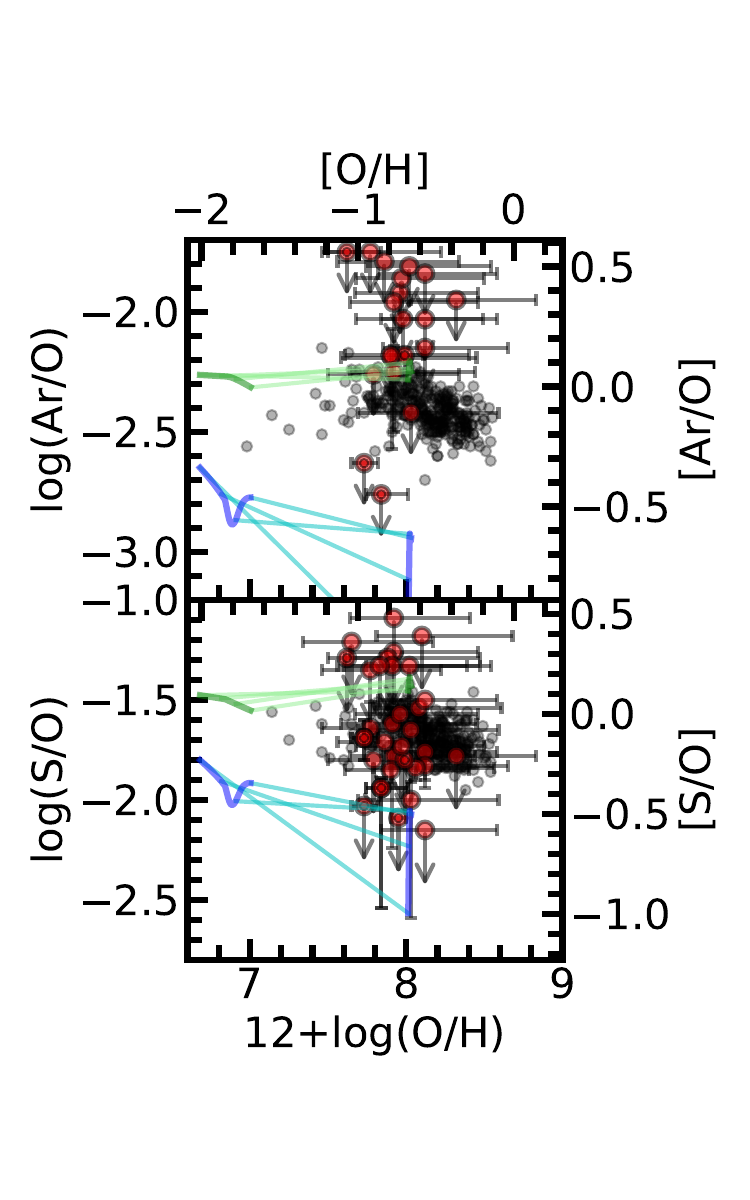}
    \caption{Ar/O and S/O as a function of metallicity. The symbols are the same as those in \tcra{Figure \ref{fig:neo}}, while we add \citet{Watanabe2023arXiv}'s chemical evolution models of hypernovae with different metallicities (green) and isochrones (light green).}
    \label{fig:ars}
\end{figure}

We obtain Ar/O and S/O ratios in the same way as the Ne/O and N/O measurements.
We derive ionic abundance ratios of Ar$^{2+}$/H$^{+}$ and S$^{+}$/H$^{+}$ from [\ariii]$\lambda$7136 and [\sii]$\lambda\lambda$6716,6731 lines respectively.
We use $T_{\rm e}$(\oii) for S$^{+}$/H$^{+}$ and $T_{\rm e}$(\siii) for Ar$^{2+}$/H$^{+}$, where $T_{\rm e}$(\siii) is calculated from $T_{\rm e}$(\oiii) and the empirical relation of \citet{Garnett1992}.
Because the ionization energy of S$^{+}$ ion is lower than that of H$^{+}$, we derive ICF(S$^{+}$) calculated until the photo-dissociation region (PDR).
Figure \ref{fig:ars} illustrates Ar/O and S/O ratios of our sample galaxies as a function of metallicity.
In addition to the CCSN models (blue), we plot \citet{Watanabe2023arXiv}'s hypernova (HN) models (green).
The observed Ar/O and S/O ratios can be explained by the combination of CCSN and HN yields (see \citealt{Watanabe2023arXiv} for more details).

%

\bibliography{library}

\begin{thebibliography}{}
\expandafter\ifx\csname natexlab\endcsname\relax\def\natexlab#1{#1}\fi
\providecommand{\url}[1]{\href{#1}{#1}}
\providecommand{\dodoi}[1]{doi:~\href{http://doi.org/#1}{\nolinkurl{#1}}}
\providecommand{\doeprint}[1]{\href{http://ascl.net/#1}{\nolinkurl{http://ascl.net/#1}}}
\providecommand{\doarXiv}[1]{\href{https://arxiv.org/abs/#1}{\nolinkurl{https://arxiv.org/abs/#1}}}

\bibitem[{{Aggarwal} \& {Keenan}(1999)}]{Aggarwal1999}
{Aggarwal}, K.~M., \& {Keenan}, F.~P. 1999, \apjs, 123, 311, \dodoi{10.1086/313232}

\bibitem[{{Arellano-C{\'o}rdova} {et~al.}(2022){Arellano-C{\'o}rdova}, {Berg}, {Chisholm}, {Arrabal Haro}, {Dickinson}, {Finkelstein}, {Leclercq}, {Rogers}, {Simons}, {Skillman}, {Trump}, \& {Kartaltepe}}]{Arellano2022}
{Arellano-C{\'o}rdova}, K.~Z., {Berg}, D.~A., {Chisholm}, J., {et~al.} 2022, \apjl, 940, L23, \dodoi{10.3847/2041-8213/ac9ab2}

\bibitem[{{Asplund} {et~al.}(2021){Asplund}, {Amarsi}, \& {Grevesse}}]{Asplund2021}
{Asplund}, M., {Amarsi}, A.~M., \& {Grevesse}, N. 2021, \aap, 653, A141, \dodoi{10.1051/0004-6361/202140445}

\bibitem[{{Astropy Collaboration} {et~al.}(2013){Astropy Collaboration}, {Robitaille}, {Tollerud}, {Greenfield}, {Droettboom}, {Bray}, {Aldcroft}, {Davis}, {Ginsburg}, {Price-Whelan}, {Kerzendorf}, {Conley}, {Crighton}, {Barbary}, {Muna}, {Ferguson}, {Grollier}, {Parikh}, {Nair}, {Unther}, {Deil}, {Woillez}, {Conseil}, {Kramer}, {Turner}, {Singer}, {Fox}, {Weaver}, {Zabalza}, {Edwards}, {Azalee Bostroem}, {Burke}, {Casey}, {Crawford}, {Dencheva}, {Ely}, {Jenness}, {Labrie}, {Lim}, {Pierfederici}, {Pontzen}, {Ptak}, {Refsdal}, {Servillat}, \& {Streicher}}]{Astropy2013}
{Astropy Collaboration}, {Robitaille}, T.~P., {Tollerud}, E.~J., {et~al.} 2013, \aap, 558, A33, \dodoi{10.1051/0004-6361/201322068}

\bibitem[{{Astropy Collaboration} {et~al.}(2022){Astropy Collaboration}, {Price-Whelan}, {Lim}, {Earl}, {Starkman}, {Bradley}, {Shupe}, {Patil}, {Corrales}, {Brasseur}, {N{\"o}the}, {Donath}, {Tollerud}, {Morris}, {Ginsburg}, {Vaher}, {Weaver}, {Tocknell}, {Jamieson}, {van Kerkwijk}, {Robitaille}, {Merry}, {Bachetti}, {G{\"u}nther}, {Aldcroft}, {Alvarado-Montes}, {Archibald}, {B{\'o}di}, {Bapat}, {Barentsen}, {Baz{\'a}n}, {Biswas}, {Boquien}, {Burke}, {Cara}, {Cara}, {Conroy}, {Conseil}, {Craig}, {Cross}, {Cruz}, {D'Eugenio}, {Dencheva}, {Devillepoix}, {Dietrich}, {Eigenbrot}, {Erben}, {Ferreira}, {Foreman-Mackey}, {Fox}, {Freij}, {Garg}, {Geda}, {Glattly}, {Gondhalekar}, {Gordon}, {Grant}, {Greenfield}, {Groener}, {Guest}, {Gurovich}, {Handberg}, {Hart}, {Hatfield-Dodds}, {Homeier}, {Hosseinzadeh}, {Jenness}, {Jones}, {Joseph}, {Kalmbach}, {Karamehmetoglu}, {Ka{\l}uszy{\'n}ski}, {Kelley}, {Kern}, {Kerzendorf}, {Koch}, {Kulumani}, {Lee}, {Ly}, {Ma}, {MacBride}, {Maljaars}, {Muna}, {Murphy}, {Norman},
  {O'Steen}, {Oman}, {Pacifici}, {Pascual}, {Pascual-Granado}, {Patil}, {Perren}, {Pickering}, {Rastogi}, {Roulston}, {Ryan}, {Rykoff}, {Sabater}, {Sakurikar}, {Salgado}, {Sanghi}, {Saunders}, {Savchenko}, {Schwardt}, {Seifert-Eckert}, {Shih}, {Jain}, {Shukla}, {Sick}, {Simpson}, {Singanamalla}, {Singer}, {Singhal}, {Sinha}, {Sip{\H{o}}cz}, {Spitler}, {Stansby}, {Streicher}, {{\v{S}}umak}, {Swinbank}, {Taranu}, {Tewary}, {Tremblay}, {de Val-Borro}, {Van Kooten}, {Vasovi{\'c}}, {Verma}, {de Miranda Cardoso}, {Williams}, {Wilson}, {Winkel}, {Wood-Vasey}, {Xue}, {Yoachim}, {Zhang}, {Zonca}, \& {Astropy Project Contributors}}]{Astropy2022}
{Astropy Collaboration}, {Price-Whelan}, A.~M., {Lim}, P.~L., {et~al.} 2022, \apj, 935, 167, \dodoi{10.3847/1538-4357/ac7c74}

\bibitem[{{Batra} \& {Baldwin}(2014)}]{Batra2014}
{Batra}, N.~D., \& {Baldwin}, J.~A. 2014, \mnras, 439, 771, \dodoi{10.1093/mnras/stu007}

\bibitem[{Berg {et~al.}(2019)Berg, Erb, Henry, Skillman, \& McQuinn}]{Berg2019}
Berg, D.~A., Erb, D.~K., Henry, R. B.~C., Skillman, E.~D., \& McQuinn, K. B.~W. 2019, ApJ, 874, 93, \dodoi{10.3847/1538-4357/ab020a}

\bibitem[{{Berg} {et~al.}(2016){Berg}, {Skillman}, {Henry}, {Erb}, \& {Carigi}}]{Berg2016}
{Berg}, D.~A., {Skillman}, E.~D., {Henry}, R. B.~C., {Erb}, D.~K., \& {Carigi}, L. 2016, \apj, 827, 126, \dodoi{10.3847/0004-637X/827/2/126}

\bibitem[{{Berrington} {et~al.}(1985){Berrington}, {Burke}, {Dufton}, \& {Kingston}}]{Berrington1985}
{Berrington}, K.~A., {Burke}, P.~G., {Dufton}, P.~L., \& {Kingston}, A.~E. 1985, Atomic Data and Nuclear Data Tables, 33, 195, \dodoi{10.1016/0092-640X(85)90001-4}

\bibitem[{{Beveridge} \& {Sneden}(1994)}]{Beveridge1994}
{Beveridge}, R.~C., \& {Sneden}, C. 1994, \aj, 108, 285, \dodoi{10.1086/117068}

\bibitem[{{Bezanson} {et~al.}(2022){Bezanson}, {Labbe}, {Whitaker}, {Leja}, {Price}, {Franx}, {Brammer}, {Marchesini}, {Zitrin}, {Wang}, {Weaver}, {Furtak}, {Atek}, {Coe}, {Cutler}, {Dayal}, {van Dokkum}, {Feldmann}, {Forster Schreiber}, {Fujimoto}, {Geha}, {Glazebrook}, {de Graaff}, {Greene}, {Juneau}, {Kassin}, {Kriek}, {Khullar}, {Maseda}, {Mowla}, {Muzzin}, {Nanayakkara}, {Nelson}, {Oesch}, {Pacifici}, {Pan}, {Papovich}, {Setton}, {Shapley}, {Smit}, {Stefanon}, {Taylor}, \& {Williams}}]{Bezanson2022arXiv}
{Bezanson}, R., {Labbe}, I., {Whitaker}, K.~E., {et~al.} 2022, arXiv e-prints, arXiv:2212.04026, \dodoi{10.48550/arXiv.2212.04026}

\bibitem[{{Blum} \& {Pradhan}(1992)}]{Blum1992}
{Blum}, R.~D., \& {Pradhan}, A.~K. 1992, \apjs, 80, 425, \dodoi{10.1086/191670}

\bibitem[{{Briley} {et~al.}(2004){Briley}, {Harbeck}, {Smith}, \& {Grebel}}]{Briley2004}
{Briley}, M.~M., {Harbeck}, D., {Smith}, G.~H., \& {Grebel}, E.~K. 2004, \aj, 127, 1588, \dodoi{10.1086/381912}

\bibitem[{{Bunker} {et~al.}(2023){Bunker}, {Saxena}, {Cameron}, {Willott}, {Curtis-Lake}, {Jakobsen}, {Carniani}, {Smit}, {Maiolino}, {Witstok}, {Curti}, {D'Eugenio}, {Jones}, {Ferruit}, {Arribas}, {Charlot}, {Chevallard}, {Giardino}, {de Graaff}, {Looser}, {Luetzgendorf}, {Maseda}, {Rawle}, {Rix}, {Rodriguez Del Pino}, {Alberts}, {Egami}, {Eisenstein}, {Endsley}, {Hainline}, {Hausen}, {Johnson}, {Rieke}, {Rieke}, {Robertson}, {Shivaei}, {Stark}, {Sun}, {Tacchella}, {Tang}, {Williams}, {Willmer}, {Baker}, {Baum}, {Bhatawdekar}, {Bowler}, {Boyett}, {Chen}, {Circosta}, {Helton}, {Ji}, {Lyu}, {Nelson}, {Parlanti}, {Perna}, {Sandles}, {Scholtz}, {Suess}, {Topping}, {Uebler}, {Wallace}, \& {Whitler}}]{Bunker2023arXiv}
{Bunker}, A.~J., {Saxena}, A., {Cameron}, A.~J., {et~al.} 2023, arXiv e-prints, arXiv:2302.07256, \dodoi{10.48550/arXiv.2302.07256}

\bibitem[{Bushouse {et~al.}(2022)Bushouse, Eisenhamer, Dencheva, Davies, Greenfield, Morrison, Hodge, Simon, Grumm, Droettboom, Slavich, Sosey, Pauly, Miller, Jedrzejewski, Hack, Davis, Crawford, Law, Gordon, Regan, Cara, MacDonald, Bradley, Shanahan, Jamieson, Teodoro, \& Williams}]{JWSTpipeline185}
Bushouse, H., Eisenhamer, J., Dencheva, N., {et~al.} 2022, JWST Calibration Pipeline, 1.8.5,  Zenodo, \dodoi{10.5281/zenodo.7429939}

\bibitem[{Calzetti {et~al.}(2000)Calzetti, Armus, Bohlin, Kinney, Koornneef, \& Storchi‐Bergmann}]{Calzetti2000}
Calzetti, D., Armus, L., Bohlin, R.~C., {et~al.} 2000, ApJ, 533, 682, \dodoi{10.1086/308692}

\bibitem[{{Cameron} {et~al.}(2023){Cameron}, {Katz}, {Rey}, \& {Saxena}}]{Cameron2023btmp}
{Cameron}, A.~J., {Katz}, H., {Rey}, M.~P., \& {Saxena}, A. 2023, \mnras, \dodoi{10.1093/mnras/stad1579}

\bibitem[{{Carretta} {et~al.}(2005){Carretta}, {Gratton}, {Lucatello}, {Bragaglia}, \& {Bonifacio}}]{Carretta2005}
{Carretta}, E., {Gratton}, R.~G., {Lucatello}, S., {Bragaglia}, A., \& {Bonifacio}, P. 2005, \aap, 433, 597, \dodoi{10.1051/0004-6361:20041892}

\bibitem[{{Charbonnel} {et~al.}(2023){Charbonnel}, {Schaerer}, {Prantzos}, {Ram{\'\i}rez-Galeano}, {Fragos}, {Kuruvanthodi}, {Marques-Chaves}, \& {Gieles}}]{Charbonnel2023}
{Charbonnel}, C., {Schaerer}, D., {Prantzos}, N., {et~al.} 2023, \aap, 673, L7, \dodoi{10.1051/0004-6361/202346410}

\bibitem[{{D'Orazi} {et~al.}(2010){D'Orazi}, {Lucatello}, {Gratton}, {Bragaglia}, {Carretta}, {Shen}, \& {Zaggia}}]{DOrazi2010}
{D'Orazi}, V., {Lucatello}, S., {Gratton}, R., {et~al.} 2010, \apjl, 713, L1, \dodoi{10.1088/2041-8205/713/1/L1}

\bibitem[{{Elvis} {et~al.}(1994){Elvis}, {Wilkes}, {McDowell}, {Green}, {Bechtold}, {Willner}, {Oey}, {Polomski}, \& {Cutri}}]{Elvis1994}
{Elvis}, M., {Wilkes}, B.~J., {McDowell}, J.~C., {et~al.} 1994, \apjs, 95, 1, \dodoi{10.1086/192093}

\bibitem[{{Ferland} {et~al.}(2013){Ferland}, {Porter}, {van Hoof}, {Williams}, {Abel}, {Lykins}, {Shaw}, {Henney}, \& {Stancil}}]{Ferland2013}
{Ferland}, G.~J., {Porter}, R.~L., {van Hoof}, P.~A.~M., {et~al.} 2013, \rmxaa, 49, 137.
\newblock \doarXiv{1302.4485}

\bibitem[{{Finkelstein} {et~al.}(2022){Finkelstein}, {Bagley}, {Ferguson}, {Wilkins}, {Kartaltepe}, {Papovich}, {Yung}, {Arrabal Haro}, {Behroozi}, {Dickinson}, {Kocevski}, {Koekemoer}, {Larson}, {Le Bail}, {Morales}, {Perez-Gonzalez}, {Burgarella}, {Dave}, {Hirschmann}, {Somerville}, {Wuyts}, {Bromm}, {Casey}, {Fontana}, {Fujimoto}, {Gardner}, {Giavalisco}, {Grazian}, {Grogin}, {Hathi}, {Hutchison}, {Jha}, {Jogee}, {Kewley}, {Kirkpatrick}, {Long}, {Lotz}, {Pentericci}, {Pierel}, {Pirzkal}, {Ravindranath}, {Ryan}, {Trump}, {Yang}, {Bhatawdekar}, {Bisigello}, {Buat}, {Calabro}, {Castellano}, {Cleri}, {Cooper}, {Croton}, {Daddi}, {Dekel}, {Elbaz}, {Franco}, {Gawiser}, {Holwerda}, {Huertas-Company}, {Jaskot}, {Leung}, {Lucas}, {Mobasher}, {Pandya}, {Tacchella}, {Weiner}, \& {Zavala}}]{Finkelstein2022arXiv}
{Finkelstein}, S.~L., {Bagley}, M.~B., {Ferguson}, H.~C., {et~al.} 2022, arXiv e-prints, arXiv:2211.05792.
\newblock \doarXiv{2211.05792}

\bibitem[{{Francis}(1993)}]{Francis1993}
{Francis}, P.~J. 1993, \apj, 407, 519, \dodoi{10.1086/172533}

\bibitem[{{Froese Fischer} \& {Tachiev}(2004)}]{FroeseFischer2004}
{Froese Fischer}, C., \& {Tachiev}, G. 2004, Atomic Data and Nuclear Data Tables, 87, 1, \dodoi{10.1016/j.adt.2004.02.001}

\bibitem[{{Galavis} {et~al.}(1998){Galavis}, {Mendoza}, \& {Zeippen}}]{Galavis1998}
{Galavis}, M.~E., {Mendoza}, C., \& {Zeippen}, C.~J. 1998, \aaps, 131, 499, \dodoi{10.1051/aas:1998435}

\bibitem[{{Garc{\'\i}a-Rojas} \& {Esteban}(2007)}]{Garcia-Rojas2007}
{Garc{\'\i}a-Rojas}, J., \& {Esteban}, C. 2007, \apj, 670, 457, \dodoi{10.1086/521871}

\bibitem[{{Garnett}(1992)}]{Garnett1992}
{Garnett}, D.~R. 1992, \aj, 103, 1330, \dodoi{10.1086/116146}

\bibitem[{{Hamann} \& {Ferland}(1993)}]{Hamman1993}
{Hamann}, F., \& {Ferland}, G. 1993, \apj, 418, 11, \dodoi{10.1086/173366}

\bibitem[{{Harikane} {et~al.}(2023{\natexlab{a}}){Harikane}, {Nakajima}, {Ouchi}, {Umeda}, {Isobe}, {Ono}, {Xu}, \& {Zhang}}]{Harikane2023carXiv}
{Harikane}, Y., {Nakajima}, K., {Ouchi}, M., {et~al.} 2023{\natexlab{a}}, arXiv e-prints, arXiv:2304.06658, \dodoi{10.48550/arXiv.2304.06658}

\bibitem[{{Harikane} {et~al.}(2023{\natexlab{b}}){Harikane}, {Ouchi}, {Oguri}, {Ono}, {Nakajima}, {Isobe}, {Umeda}, {Mawatari}, \& {Zhang}}]{Harikane2023a}
{Harikane}, Y., {Ouchi}, M., {Oguri}, M., {et~al.} 2023{\natexlab{b}}, \apjs, 265, 5, \dodoi{10.3847/1538-4365/acaaa9}

\bibitem[{Heger {et~al.}(2003)Heger, Fryer, Woosley, Langer, \& Hartmann}]{Heger2003}
Heger, A., Fryer, C.~L., Woosley, S.~E., Langer, N., \& Hartmann, D.~H. 2003, ApJ, 591, 288, \dodoi{10.1086/375341}

\bibitem[{{Isobe} {et~al.}(2023){Isobe}, {Ouchi}, {Nakajima}, {Harikane}, {Ono}, {Xu}, {Zhang}, \& {Umeda}}]{Isobe2023}
{Isobe}, Y., {Ouchi}, M., {Nakajima}, K., {et~al.} 2023, arXiv e-prints, arXiv:2301.06811, \dodoi{10.48550/arXiv.2301.06811}

\bibitem[{{Isobe} {et~al.}(2022){Isobe}, {Ouchi}, {Suzuki}, {Moriya}, {Nakajima}, {Nomoto}, {Rauch}, {Harikane}, {Kojima}, {Ono}, {Fujimoto}, {Inoue}, {Kim}, {Komiyama}, {Kusakabe}, {Lee}, {Maseda}, {Matthee}, {Michel-Dansac}, {Nagao}, {Nanayakkara}, {Nishigaki}, {Onodera}, {Sugahara}, \& {Xu}}]{Isobe2022}
{Isobe}, Y., {Ouchi}, M., {Suzuki}, A., {et~al.} 2022, \apj, 925, 111, \dodoi{10.3847/1538-4357/ac3509}

\bibitem[{Izotov {et~al.}(2006)Izotov, Stasi{\'{n}}ska, Meynet, Guseva, \& Thuan}]{Izotov2006}
Izotov, Y.~I., Stasi{\'{n}}ska, G., Meynet, G., Guseva, N.~G., \& Thuan, T.~X. 2006, A{\&}A, 448, 955, \dodoi{10.1051/0004-6361:20053763}

\bibitem[{{Jones} {et~al.}(2023){Jones}, {Sanders}, {Chen}, {Wang}, {Morishita}, {Roberts-Borsani}, {Treu}, {Dressler}, {Merlin}, {Paris}, {Santini}, {Bergamini}, {Huntzinger}, {Nanayakkara}, {Boyett}, {Bradac}, {Brammer}, {Calabro}, {Glazebrook}, {Grasha}, {Mascia}, {Pentericci}, {Trenti}, \& {Vulcani}}]{Jones2023arXiv}
{Jones}, T., {Sanders}, R., {Chen}, Y., {et~al.} 2023, arXiv e-prints, arXiv:2301.07126, \dodoi{10.48550/arXiv.2301.07126}

\bibitem[{{Kisielius} {et~al.}(2009){Kisielius}, {Storey}, {Ferland}, \& {Keenan}}]{Kisielius2009}
{Kisielius}, R., {Storey}, P.~J., {Ferland}, G.~J., \& {Keenan}, F.~P. 2009, \mnras, 397, 903, \dodoi{10.1111/j.1365-2966.2009.14989.x}

\bibitem[{{Kojima} {et~al.}(2017){Kojima}, {Ouchi}, {Nakajima}, {Shibuya}, {Harikane}, \& {Ono}}]{Kojima2017}
{Kojima}, T., {Ouchi}, M., {Nakajima}, K., {et~al.} 2017, \pasj, 69, 44, \dodoi{10.1093/pasj/psx017}

\bibitem[{{Kojima} {et~al.}(2021){Kojima}, {Ouchi}, {Rauch}, {Ono}, {Nakajima}, {Isobe}, {Fujimoto}, {Harikane}, {Hashimoto}, {Hayashi}, {Komiyama}, {Kusakabe}, {Kim}, {Lee}, {Mukae}, {Nagao}, {Onodera}, {Shibuya}, {Sugahara}, {Umemura}, \& {Yabe}}]{Kojima2021}
{Kojima}, T., {Ouchi}, M., {Rauch}, M., {et~al.} 2021, \apj, 913, 22, \dodoi{10.3847/1538-4357/abec3d}

\bibitem[{{Kroupa}(2001)}]{Kroupa2001}
{Kroupa}, P. 2001, \mnras, 322, 231, \dodoi{10.1046/j.1365-8711.2001.04022.x}

\bibitem[{{Larson} {et~al.}(2023){Larson}, {Finkelstein}, {Kocevski}, {Hutchison}, {Trump}, {Arrabal Haro}, {Bromm}, {Cleri}, {Dickinson}, {Fujimoto}, {Kartaltepe}, {Koekemoer}, {Papovich}, {Pirzkal}, {Tacchella}, {Zavala}, {Bagley}, {Behroozi}, {Champagne}, {Cole}, {Jung}, {Morales}, {Yang}, {Zhang}, {Zitrin}, {Amor{\'\i}n}, {Burgarella}, {Casey}, {Ch{\'a}vez Ortiz}, {Cox}, {Chworowsky}, {Fontana}, {Gawiser}, {Grazian}, {Grogin}, {Harish}, {Hathi}, {Hirschmann}, {Holwerda}, {Juneau}, {Leung}, {Lucas}, {McGrath}, {P{\'e}rez-Gonz{\'a}lez}, {Rigby}, {Seill{\'e}}, {Simons}, {Weiner}, {Wilkins}, {Yung}, \& {The CEERS Team}}]{Larson2023arXiv}
{Larson}, R.~L., {Finkelstein}, S.~L., {Kocevski}, D.~D., {et~al.} 2023, arXiv e-prints, arXiv:2303.08918, \dodoi{10.48550/arXiv.2303.08918}

\bibitem[{{Limongi} \& {Chieffi}(2018)}]{Limongi2018}
{Limongi}, M., \& {Chieffi}, A. 2018, \apjs, 237, 13, \dodoi{10.3847/1538-4365/aacb24}

\bibitem[{{Llerena} {et~al.}(2022){Llerena}, {Amor{\'\i}n}, {Cullen}, {Pentericci}, {Calabr{\`o}}, {McLure}, {Carnall}, {P{\'e}rez-Montero}, {Marchi}, {Bongiorno}, {Castellano}, {Fontana}, {McLeod}, {Talia}, {Hathi}, {Hibon}, {Mannucci}, {Saxena}, {Schaerer}, \& {Zamorani}}]{Llerena2022}
{Llerena}, M., {Amor{\'\i}n}, R., {Cullen}, F., {et~al.} 2022, \aap, 659, A16, \dodoi{10.1051/0004-6361/202141651}

\bibitem[{{Llerena} {et~al.}(2023){Llerena}, {Amor{\'\i}n}, {Pentericci}, {Calabr{\`o}}, {Shapley}, {Boutsia}, {P{\'e}rez-Montero}, {V{\'\i}lchez}, \& {Nakajima}}]{Llerena2023}
{Llerena}, M., {Amor{\'\i}n}, R., {Pentericci}, L., {et~al.} 2023, \aap, 676, A53, \dodoi{10.1051/0004-6361/202346232}

\bibitem[{{Luridiana} {et~al.}(2015){Luridiana}, {Morisset}, \& {Shaw}}]{Luridiana2015}
{Luridiana}, V., {Morisset}, C., \& {Shaw}, R.~A. 2015, \aap, 573, A42, \dodoi{10.1051/0004-6361/201323152}

\bibitem[{{Maeder} {et~al.}(2015){Maeder}, {Meynet}, \& {Chiappini}}]{Maeder2015}
{Maeder}, A., {Meynet}, G., \& {Chiappini}, C. 2015, \aap, 576, A56, \dodoi{10.1051/0004-6361/201424153}

\bibitem[{{Maiolino} {et~al.}(2023){Maiolino}, {Scholtz}, {Witstok}, {Carniani}, {D'Eugenio}, {de Graaff}, {Uebler}, {Tacchella}, {Curtis-Lake}, {Arribas}, {Bunker}, {Charlot}, {Chevallard}, {Curti}, {Looser}, {Maseda}, {Rawle}, {Rodriguez Del Pino}, {Willott}, {Egami}, {Eisenstein}, {Hainline}, {Robertson}, {Williams}, {Willmer}, {Baker}, {Boyett}, {DeCoursey}, {Fabian}, {Helton}, {Ji}, {Jones}, {Kumari}, {Laporte}, {Nelson}, {Perna}, {Sandles}, {Shivaei}, \& {Sun}}]{Maiolino2023arXiv}
{Maiolino}, R., {Scholtz}, J., {Witstok}, J., {et~al.} 2023, arXiv e-prints, arXiv:2305.12492, \dodoi{10.48550/arXiv.2305.12492}

\bibitem[{{Marques-Chaves} {et~al.}(2023){Marques-Chaves}, {Schaerer}, {Kuruvanthodi}, {Korber}, {Prantzos}, {Charbonnel}, {Weibel}, {Izotov}, {Messa}, {Brammer}, {Dessauges-Zavadsky}, \& {Oesch}}]{Marques-Chaves2023arXiv}
{Marques-Chaves}, R., {Schaerer}, D., {Kuruvanthodi}, A., {et~al.} 2023, arXiv e-prints, arXiv:2307.04234, \dodoi{10.48550/arXiv.2307.04234}

\bibitem[{{McLaughlin} \& {Bell}(2000)}]{McLaughlin2000}
{McLaughlin}, B.~M., \& {Bell}, K.~L. 2000, JPhB, 33, 597, \dodoi{10.1088/0953-4075/33/4/301}

\bibitem[{{Morisset} {et~al.}(2015){Morisset}, {Delgado-Inglada}, \& {Flores-Fajardo}}]{Morisset2015}
{Morisset}, C., {Delgado-Inglada}, G., \& {Flores-Fajardo}, N. 2015, \rmxaa, 51, 103, \dodoi{10.48550/arXiv.1412.5349}

\bibitem[{{Munoz Burgos} {et~al.}(2009){Munoz Burgos}, {Loch}, {Ballance}, \& {Boivin}}]{MunozBurgos2009}
{Munoz Burgos}, J.~M., {Loch}, S.~D., {Ballance}, C.~P., \& {Boivin}, R.~F. 2009, \aap, 500, 1253, \dodoi{10.1051/0004-6361/200911743}

\bibitem[{{Nagele} \& {Umeda}(2023)}]{Nagele2023}
{Nagele}, C., \& {Umeda}, H. 2023, \apjl, 949, L16, \dodoi{10.3847/2041-8213/acd550}

\bibitem[{{Nakajima} {et~al.}(2023){Nakajima}, {Ouchi}, {Isobe}, {Harikane}, {Zhang}, {Ono}, {Umeda}, \& {Oguri}}]{Nakajima2023}
{Nakajima}, K., {Ouchi}, M., {Isobe}, Y., {et~al.} 2023, arXiv e-prints, arXiv:2301.12825, \dodoi{10.48550/arXiv.2301.12825}

\bibitem[{{Nicholls} {et~al.}(2017){Nicholls}, {Sutherland}, {Dopita}, {Kewley}, \& {Groves}}]{Nicholls2017}
{Nicholls}, D.~C., {Sutherland}, R.~S., {Dopita}, M.~A., {Kewley}, L.~J., \& {Groves}, B.~A. 2017, \mnras, 466, 4403, \dodoi{10.1093/mnras/stw3235}

\bibitem[{{Nomoto} {et~al.}(2013){Nomoto}, {Kobayashi}, \& {Tominaga}}]{Nomoto2013}
{Nomoto}, K., {Kobayashi}, C., \& {Tominaga}, N. 2013, \araa, 51, 457, \dodoi{10.1146/annurev-astro-082812-140956}

\bibitem[{{Norris} {et~al.}(2013){Norris}, {Yong}, {Bessell}, {Christlieb}, {Asplund}, {Gilmore}, {Wyse}, {Beers}, {Barklem}, {Frebel}, \& {Ryan}}]{Norris2013}
{Norris}, J.~E., {Yong}, D., {Bessell}, M.~S., {et~al.} 2013, \apj, 762, 28, \dodoi{10.1088/0004-637X/762/1/28}

\bibitem[{Oesch {et~al.}(2016)Oesch, Brammer, van Dokkum, Illingworth, Bouwens, Labb{\'{e}}, Franx, Momcheva, Ashby, Fazio, Gonzalez, Holden, Magee, Skelton, Smit, Spitler, Trenti, \& Willner}]{Oesch2016}
Oesch, P.~A., Brammer, G., van Dokkum, P.~G., {et~al.} 2016, ApJ, 819, 129, \dodoi{10.3847/0004-637x/819/2/129}

\bibitem[{{Pilyugin} {et~al.}(2012){Pilyugin}, {V{\'\i}lchez}, {Mattsson}, \& {Thuan}}]{Pilyugin2012}
{Pilyugin}, L.~S., {V{\'\i}lchez}, J.~M., {Mattsson}, L., \& {Thuan}, T.~X. 2012, \mnras, 421, 1624, \dodoi{10.1111/j.1365-2966.2012.20420.x}

\bibitem[{{Pontoppidan} {et~al.}(2022){Pontoppidan}, {Barrientes}, {Blome}, {Braun}, {Brown}, {Carruthers}, {Coe}, {DePasquale}, {Espinoza}, {Marin}, {Gordon}, {Henry}, {Hustak}, {James}, {Jenkins}, {Koekemoer}, {LaMassa}, {Law}, {Lockwood}, {Moro-Martin}, {Mullally}, {Pagan}, {Player}, {Proffitt}, {Pulliam}, {Ramsay}, {Ravindranath}, {Reid}, {Robberto}, {Sabbi}, {Ubeda}, {Balogh}, {Flanagan}, {Gardner}, {Hasan}, {Meinke}, \& {Nota}}]{Pontoppidan2022}
{Pontoppidan}, K.~M., {Barrientes}, J., {Blome}, C., {et~al.} 2022, \apjl, 936, L14, \dodoi{10.3847/2041-8213/ac8a4e}

\bibitem[{{Portinari} {et~al.}(1998){Portinari}, {Chiosi}, \& {Bressan}}]{Portinari1998}
{Portinari}, L., {Chiosi}, C., \& {Bressan}, A. 1998, \aap, 334, 505.
\newblock \doarXiv{astro-ph/9711337}

\bibitem[{{Rynkun} {et~al.}(2019){Rynkun}, {Gaigalas}, \& {J{\"o}nsson}}]{Rynkun2019}
{Rynkun}, P., {Gaigalas}, G., \& {J{\"o}nsson}, P. 2019, \aap, 623, A155, \dodoi{10.1051/0004-6361/201834931}

\bibitem[{{Salpeter}(1955)}]{Salpeter1955}
{Salpeter}, E.~E. 1955, \apj, 121, 161, \dodoi{10.1086/145971}

\bibitem[{{Senchyna} {et~al.}(2023){Senchyna}, {Plat}, {Stark}, \& {Rudie}}]{Senchyna2023arXiv}
{Senchyna}, P., {Plat}, A., {Stark}, D.~P., \& {Rudie}, G.~C. 2023, arXiv e-prints, arXiv:2303.04179, \dodoi{10.48550/arXiv.2303.04179}

\bibitem[{{Stanway} \& {Eldridge}(2018)}]{Stanway2018}
{Stanway}, E.~R., \& {Eldridge}, J.~J. 2018, \mnras, 479, 75, \dodoi{10.1093/mnras/sty1353}

\bibitem[{{Steidel} {et~al.}(2016){Steidel}, {Strom}, {Pettini}, {Rudie}, {Reddy}, \& {Trainor}}]{Steidel2016}
{Steidel}, C.~C., {Strom}, A.~L., {Pettini}, M., {et~al.} 2016, \apj, 826, 159, \dodoi{10.3847/0004-637X/826/2/159}

\bibitem[{{Storey} \& {Hummer}(1995)}]{Storey1995}
{Storey}, P.~J., \& {Hummer}, D.~G. 1995, \mnras, 272, 41, \dodoi{10.1093/mnras/272.1.41}

\bibitem[{{Storey} {et~al.}(2014){Storey}, {Sochi}, \& {Badnell}}]{Storey2014}
{Storey}, P.~J., {Sochi}, T., \& {Badnell}, N.~R. 2014, \mnras, 441, 3028, \dodoi{10.1093/mnras/stu777}

\bibitem[{Suzuki \& Maeda(2018)}]{Suzuki2018}
Suzuki, A., \& Maeda, K. 2018, MNRAS, 478, 110, \dodoi{10.1093/mnras/sty999}

\bibitem[{{Tayal} \& {Zatsarinny}(2010)}]{Tayal2010}
{Tayal}, S.~S., \& {Zatsarinny}, O. 2010, \apjs, 188, 32, \dodoi{10.1088/0067-0049/188/1/32}

\bibitem[{{Telles} {et~al.}(2014){Telles}, {Thuan}, {Izotov}, \& {Carrasco}}]{Telles2014}
{Telles}, E., {Thuan}, T.~X., {Izotov}, Y.~I., \& {Carrasco}, E.~R. 2014, \aap, 561, A64, \dodoi{10.1051/0004-6361/201219270}

\bibitem[{{The Astropy Collaboration}(2018)}]{Astropy2018}
{The Astropy Collaboration}. 2018, {astropy v3.1: a core python package for astronomy}, 3.1, Zenodo,  Zenodo, \dodoi{10.5281/zenodo.4080996}

\bibitem[{Tominaga {et~al.}(2007)Tominaga, Umeda, \& Nomoto}]{Tominaga2007}
Tominaga, N., Umeda, H., \& Nomoto, K. 2007, ApJ, 660, 516, \dodoi{10.1086/513063}

\bibitem[{{Treu} {et~al.}(2022){Treu}, {Roberts-Borsani}, {Bradac}, {Brammer}, {Fontana}, {Henry}, {Mason}, {Morishita}, {Pentericci}, {Wang}, {Acebron}, {Bagley}, {Bergamini}, {Belfiori}, {Bonchi}, {Boyett}, {Boutsia}, {Calabr{\'o}}, {Caminha}, {Castellano}, {Dressler}, {Glazebrook}, {Grillo}, {Jacobs}, {Jones}, {Kelly}, {Leethochawalit}, {Malkan}, {Marchesini}, {Mascia}, {Mercurio}, {Merlin}, {Nanayakkara}, {Nonino}, {Paris}, {Poggianti}, {Rosati}, {Santini}, {Scarlata}, {Shipley}, {Strait}, {Trenti}, {Tubthong}, {Vanzella}, {Vulcani}, \& {Yang}}]{Treu2022}
{Treu}, T., {Roberts-Borsani}, G., {Bradac}, M., {et~al.} 2022, \apj, 935, 110, \dodoi{10.3847/1538-4357/ac8158}

\bibitem[{Vincenzo {et~al.}(2016)Vincenzo, Belfiore, Maiolino, Matteucci, \& Ventura}]{Vincenzo2016}
Vincenzo, F., Belfiore, F., Maiolino, R., Matteucci, F., \& Ventura, P. 2016, MNRAS, 458, 3466, \dodoi{10.1093/mnras/stw532}

\bibitem[{{Watanabe} {et~al.}(2023){Watanabe}, {Ouchi}, {Nakajima}, {Isobe}, {Tominaga}, {Suzuki}, {Ishigaki}, {Nomoto}, {Takahashi}, {Harikane}, {Hatano}, {Kusakabe}, {Moriya}, {Nishigaki}, {Ono}, {Onodera}, \& {Sugahara}}]{Watanabe2023arXiv}
{Watanabe}, K., {Ouchi}, M., {Nakajima}, K., {et~al.} 2023, arXiv e-prints, arXiv:2305.02078, \dodoi{10.48550/arXiv.2305.02078}

\bibitem[{{Watanabe et al.}(in prep.)}]{Watanabeprep}
{Watanabe et al.}, K. in prep.

\bibitem[{{Wiese} {et~al.}(1996){Wiese}, {Fuhr}, \& {Deters}}]{Wiese1996}
{Wiese}, W.~L., {Fuhr}, J.~R., \& {Deters}, T.~M. 1996, {Atomic transition probabilities of carbon, nitrogen, and oxygen : a critical data compilation}

\bibitem[{{Woosley} \& {Janka}(2005)}]{Woosley2005}
{Woosley}, S., \& {Janka}, T. 2005, Nature Physics, 1, 147, \dodoi{10.1038/nphys172}

\bibitem[{{Yajima} {et~al.}(2022{\natexlab{a}}){Yajima}, {Abe}, {Fukushima}, {Ono}, {Harikane}, {Ouchi}, {Hashimoto}, \& {Khochfar}}]{Yajima2022barXiv}
{Yajima}, H., {Abe}, M., {Fukushima}, H., {et~al.} 2022{\natexlab{a}}, arXiv e-prints, arXiv:2211.12970, \dodoi{10.48550/arXiv.2211.12970}

\bibitem[{{Yajima} {et~al.}(2022{\natexlab{b}}){Yajima}, {Abe}, {Khochfar}, {Nagamine}, {Inoue}, {Kodama}, {Arata}, {Dalla Vecchia}, {Fukushima}, {Hashimoto}, {Kashikawa}, {Kubo}, {Li}, {Matsuda}, {Mawatari}, {Ouchi}, \& {Umehata}}]{Yajima2022}
{Yajima}, H., {Abe}, M., {Khochfar}, S., {et~al.} 2022{\natexlab{b}}, \mnras, 509, 4037, \dodoi{10.1093/mnras/stab3092}

\bibitem[{{Zamorani} {et~al.}(1981){Zamorani}, {Henry}, {Maccacaro}, {Tananbaum}, {Soltan}, {Avni}, {Liebert}, {Stocke}, {Strittmatter}, {Weymann}, {Smith}, \& {Condon}}]{Zamorani1981}
{Zamorani}, G., {Henry}, J.~P., {Maccacaro}, T., {et~al.} 1981, \apj, 245, 357, \dodoi{10.1086/158815}

\end{thebibliography}


\end{document}